\documentstyle[epsf]{article}
\def\z0{\rm Z^0}
\def\as{\alpha_{\rm s}}
\newcommand{\oa}{{\cal O}(\as)}
\newcommand{\oaa}{{\cal O}(\as^2)}
\newcommand{\oaaa}{{\cal O}(\as^3)}
\newcommand{\epem}{{\rm e^+\rm e^-}}
\newcommand{\yc}{y_{\rm cut}}
\newcommand{\amz}{\as(M_{\rm Z^0})}

\def\mz{M_{\rm Z^0}}
\def\xmu{x_{\mu}}
\def\rz{R_{\rm Z}}
\def\d2{D_2}
\def\oq{\char'134}
\def\lamsb{\Lambda_{\overline{\mbox{\scriptsize MS}}}}
\def\ecm{E_{cm}}

\def\m2{\mu^2}
\def\q{\rm q}

\def\p{\rm p}

\def\q2{Q^2}
\def\asq{\as (\q2 )}
\def\wamz{\overline{\as}(M_{\rm Z^0})}
\def\asb{\hat{\alpha}_s}
\def\dwas{\Delta\overline{\as}}
\def\R{\cal{R}}
\def\msbar{\overline{\mbox{MS}}}
\def\F{{\rm F}}

\begin{document}
\hyphenation{extra-po-lated hadro-ni-za-tion}

%
%

\title{DETERMINATION OF THE QCD COUPLING $\as$}

\author{S. BETHKE \\
Max Planck Institut f\"ur Physik\\ 
80805 M\"unchen, Germany \\
{\small e-mail: bethke@mppmu.mpg.de}} 

\date{}

\maketitle

\begin{abstract} 
Theoretical basics and experimental determinations of the coupling parameter of
the Strong Interaction, $\as$, are reviewed. 
The world average value of $\as$, expressed at the energy scale of the rest mass
of the $\z0$ boson, is determined from
analyses which are based on complete NNLO perturbative QCD.
The result is
$\amz = 0.1184 \pm 0.0031$. 
No significant deviations or systematic biases of
subsamples of experimental results are found.
From the observed energy dependence of $\as$, which is in excellent agreement
with the expectations of QCD, the number of colour degrees of freedom can be
constrained to $N_c = 3.03 \pm 0.12$.  

\end{abstract}

\vspace*{-10.5cm}
\begin{flushright}
MPI-PhE/2000-07 \\
April 2000
\end{flushright}
\vspace*{8.9cm}

\section{Preface} \label{sec:preface}


The coupling strength $\as$ is the basic free parameter of Quantum
Chromodynamics (QCD), the theory of the Strong Interaction \cite{qcd}
which is one of the four fundamental forces of nature.
QCD describes the interaction of quarks through the exchange of an octet of
massless vector gauge bosons, the gluons, using similar concepts as known from
Quantum Electrodynamics, QED. 
QCD, however, is more complex than QED because
quarks and gluons, the analogues to electrons and photons in QED, are not
observed as free particles but are confined inside hadrons.

Confinement implies that the coupling strength $\as$, the analogue to the fine
structure constant $\alpha$ in QED, becomes large 
in the regime of large-distance or low-momentum transfer 
interactions\footnote{
In the world of quantum physics, \oq large" distances $\Delta s$ correspond to
$\Delta s >$~1~fm, \oq low" momentum transfers to $Q < $~1~GeV/c...}.
%
Conversely, quarks and gluons are probed to behave like free particles, for
short time intervals\footnote{
... and \oq short" time intervals correspond to $\Delta t \ < \ 10 ^{-24}$~s.},
in high-energy or short-distance
reactions; they are said to
be \oq asymptotically free", i.e. $\as \ \rightarrow$~0 for momentum transfers 
$Q\ \rightarrow \ \infty$.

Within QCD, the phenomenology of confinement and of asymptotic freedom is
realized by introducing a new quantum number, called \oq colour charge".
Quarks carry one out of three different colour charges, while hadrons are
colourless bound states of 3 quarks or 3 antiquarks (\oq baryons"), or of a quark
and an anti-quark (\oq mesons"). 
Gluons, in contrast to photons which do not carry
(electrical) charge by themselves, have two colour charges.
This concept leads to 
the process of gluon self-interaction, which in turn,
through the effect of gluon vacuum polarization,
gives rise to asymptotic
freedom, i.e. the decrease of $\as$ with increasing momentum transfer.

As in the case of QED, QCD predicts the {\it energy dependence}
of $\as$, while the actual value of $\as$, at a given
energy or four momentum transfer scale\footnote{Here and in the following, a
system of units is utilized where the speed of light and Planck's constant are put
to unity, $c = \hbar = 1$, such that energies, momenta and masses are all given in
units of GeV.}
$Q$, is not predicted but must be
determined from experiment.

Determining $\as$ at a specific energy scale $Q$ is therefore a
fundamental measurement, to be compared with measurements of 
the electromagnetic coupling $\alpha$, of the
elementary electric charge, or of the gravitational constant.
{\em Testing} QCD as such, however, requires the measurement of $\as$ at least at
{\em two different} energy scales, and/or at different processes: one measurement
fixes the free parameter and thus provides accurate predictions for the value of
$\as$ at other energy scales and/or at other processes.

\begin{figure}[ht]
\begin{center}
\epsfxsize13.5cm\epsffile{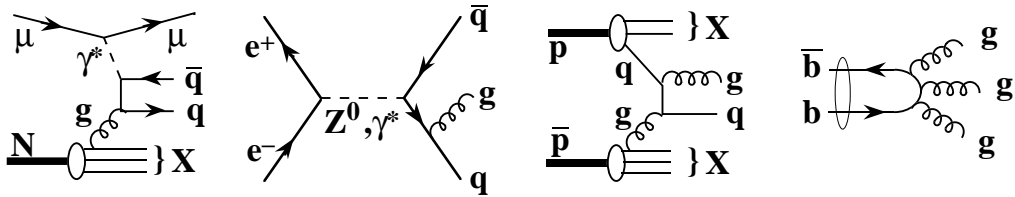} 
\end{center}
\caption{Examples of
Feynman diagrams describing hadronic final states in processes which are used
to measure $\as$.
\label{fig:qcd-feynmans}}
\end{figure}

In general, $\as$ can be determined in dynamic particle reactions
involving in- or outgoing quarks and gluons, which manifest themselves 
as hadrons. 
Examples of Feynman diagrams describing hadronic final states in deep inelastic
lepton-nucleon scattering (DIS), electron-positron annihilation ($\epem$), hadron
collisions and quarkonia decays
are shown in
figure~\ref{fig:qcd-feynmans}.

In this report, at the turn of the millennium, the current status of
measurements of $\as$ is reviewed. 
Theoretical basics of $\as$ and QCD are given in Section~\ref{sec:theory}.
Measurements of $\as$ from deep inelastic scattering, from $\epem$ annihilation
processes, from hadron colliders and from heavy quarkonia decays are
discussed in Sections 3 to 6, respectively.
A global summary of these results, a determination 
of the world average value of $\amz$ and quantitative studies of the energy
dependence of $\as$ are presented in Section~\ref{sec:summary}.
Section~\ref{sec:conclusion} concludes and gives an outlook to
future developments.

\section{QCD and $\as$: basic theoretical predictions} \label{sec:theory}
 
The concepts of QCD are described in many text books and review articles, see
e.g. references \cite{ellis-book,yndurain-book}. 
A brief review of the basics of perturbative QCD
and of the coupling strength $\as$,
like the concepts of renormalization, asymptotic freedom and confinement, the
$\Lambda$ parameter, the treatment of quark masses and thresholds,
perturbative predictions of physical observables and
renormalization scale dependence, and of nonperturbative methods like lattice
calculations
will be presented in the following subsections. 

\subsection{Renormalization}  

In quantum field theories like QCD and QED, dimensionless physical
quantities $\cal{R}$ can be expressed by a perturbation series in powers of the
coupling parameter $\as$ or $\alpha$, respectively.
Consider $\cal{R}$ depending on $\as$ and on a single energy scale $Q$.
This scale
shall be larger than any other relevant, dimensionful parameter such as quark
masses. 
In the following, these masses are therefore set to zero.

When calculating $\cal{R}$ as a perturbation series in $\as$, ultraviolet
divergencies occur.
Because $\cal{R}$ must retain physical values, these divergencies are
removed by a procedure called \oq renormalization".
This
introduces a second mass or energy scale, $\mu$, which
represents the point at which the subtraction to remove the
ultraviolet divergencies is actually performed.
As a consequence of this procedure, $\cal{R}$ and $\as$ become functions of
the renormalization scale $\mu$.
Since $\cal{R}$ is  dimensionless, we assume that it only depends on the ratio
$Q^2 / \mu^2 $ and on the renormalized coupling $\as (\mu^2 )$:
$$ {\cal R} \equiv {\cal R}(Q^2 / \mu^2, \as );\ \as \equiv \as (\mu^2). $$

Because the choice of $\mu$ is arbitrary, however, $\cal{R}$ cannot depend
on $\mu$, for a fixed value of the coupling, such that

\begin{equation} \label{eq-muindependence}
 \mu^2 \frac{{\rm d}}{{\rm d} \mu^2} {\cal R} (Q^2 / \mu^2 , \as )
= \left( \mu^2 \frac{\partial }{\partial \mu^2 } + \mu^2 \frac{\partial
\as}{\partial \mu^2} \frac{\partial }{\partial \as } \right) {\cal R} 
=^{\hskip -5pt !}\ 0 \ ,
\end{equation}

\noindent where the convention of multiplying the whole equation with $\mu^2$ is
applied in order to keep the expression dimensionless. 
Equation~\ref{eq-muindependence} implies that any explicit dependence of
$\cal{R}$ on
$\mu$ must be cancelled by an appropriate $\mu$-dependence of $\as$.
It would therefore be natural to identify the renormalization scale with the
physical energy scale of the process, $\mu^2 = Q^2$, eliminating the
uncomfortable presence of a second and unspecified scale. 
In this case, $\as$ transforms to the \oq running coupling constant"
$\asq$, and the energy dependence of $\cal{R}$ enters only 
through the energy dependence of $\asq$.

\subsection{$\as$ and its energy dependence}

While QCD does not predict the actual size of $\as$ at a particular 
energy scale, its energy dependence 
is precisely determined.
If the renormalized coupling $\as (\mu^2)$ can be fixed (i.e. measured) at a given
scale $\mu^2$, QCD definitely predicts  the size of $\as$ at
any other energy scale $Q^2$ through the renormalization group equation
\begin{equation} \label{eq-rge}
Q^2 \frac{\partial \asq}{\partial Q^2} = \beta \left( \asq \right) \ .
\end{equation}
\noindent The perturbative expansion of the $\beta$ function, including higher
order loop corrections to the bare vertices of the theory, is calculated to
complete 4-loop approximation \cite{beta4loop}:
\begin{equation} \label{eq-betafunction}
\beta (\asq ) = - \beta_0 \as^2(Q^2) - \beta_1 \as^3(Q^2) - \beta_2 \as^4(Q^2) -
\beta_3
\as^5(Q^2)  + {\cal O}(\as^6)\ ,
\end{equation}
\noindent where
\begin{eqnarray} \label{eq-betas}
\beta_0 &=& \frac{33 - 2 N_f}{12 \pi}\ , \nonumber \\
\beta_1 &=& \frac{153 - 19 N_f}{24 \pi^2}\ , \nonumber \\
\beta_2 &=& \frac{77139 - 15099 N_f + 325 N_f^2}{3456 \pi^3}\ , \nonumber \\
\beta_3 &\approx & \frac{29243 - 6946.3 N_f + 405.089 N_f^2 + 1.49931 N_f^3}
        {256 \pi^4} \ ,
\end{eqnarray}
\noindent and $N_f$ is the number of active quark flavours at the energy
scale $Q$.
The numerical constants in equation~\ref{eq-betas} are functions of the group
constants
$C_A$ and $C_F$, which for QCD --- exhibiting $SU(3)$ symmetry --- have values of
$C_A = 3$ and $C_F = 4/3$ (see e.g. \cite{ellis-book} for more details).
$\beta_0$ and $\beta_1$ are independent of the
renormalization scheme, while all higher order $\beta$ coefficients are
scheme dependent.

\subsection{Asymptotic freedom and confinement}

A solution of equation~\ref{eq-betafunction} in 1-loop approximation, i.e.
neglecting
$\beta_1$ and higher order terms, is
\begin{equation} \label{eq-as1loop}
\asq = \frac{\as (\mu^2 )}{1 + \as (\mu^2 ) \beta_0 \ln{\frac{\q2}{\mu^2}}}\ .
\end{equation}
\noindent 
Apart from giving a relation between $\asq$ and $\as (\mu^2 )$,
equation~\ref{eq-as1loop} also demonstrates the property of asymptotic freedom:
if $\q2$ becomes large and $\beta_0$ is positive, i.e. if
$N_f < 17$, $\asq$ will decrease to zero.

Likewise, equation~\ref{eq-as1loop} indicates that $\asq$ grows to large values and
actually diverges to infinity at small $Q^2$: for instance, with
$\as ( \mu^2 \equiv M^2_{\rm Z^0}) = 0.12$ and for 
typical values of $N_f = 2\ ...\ 5$,
$\asq$ exceeds unity for $Q^2 \leq \cal{O} \rm{(100~MeV~...~1~GeV)}$.
Clearly, this is the region where perturbative expansions in $\as$ are not
meaningful anymore, and we may regard energy scales of $\mu^2$ and $Q^2$ below
the order of 1~GeV as the nonperturbative region where confinement sets in, and
where equations~\ref{eq-betafunction} and~\ref{eq-as1loop} cannot be applied.

Including $\beta_1$ and higher order terms, similar but more complicated
relations for $\asq$, as a function of $\as (\mu^2 )$ and of
$\ln{\frac{\q2}{\mu^2}}$ as in equation~\ref{eq-as1loop}, emerge.
They can be solved numerically, such that for a given value of $\as (\mu^2 )$, 
choosing a suitable reference scale like the mass of the $\z0$ boson,
$\mu = M_{\z0}$,
$\asq$ can be accurately determined at any energy scale $\q2 \geq 1~\rm{GeV}^2$.

\subsection{The $\Lambda$ parameter}

Alternatively, as another parametrization of 
$\asq$, a parameter called $\Lambda$ is introduced as a constant of integration
of the $\beta$-function, such that, in leading order,
\begin{equation} \label{eq-as1loop-2}
\asq = \frac{1}{\beta_0 \ln (\q2 / \Lambda^2)}\ .
\end{equation}
\noindent 
Equations~\ref{eq-as1loop} and~\ref{eq-as1loop-2} are equivalent with each other
if
$$\Lambda^2 = \frac{\mu^2}{e^{1/\left( \beta_0 \as (\mu^2)\right) }}\ .$$
\noindent
Hence, $\as (\mu^2)$ is replaced by a suitable choice of the $\Lambda$ parameter,
which technically is identical to the energy scale $Q$ where $\asq$ diverges to
infinity, $\asq \rightarrow \infty$ for $Q^2 \rightarrow \Lambda^2$.
To give a numerical example, $\Lambda \approx 0.1$~GeV for 
$\as (\mz \equiv 91.2\ \rm{GeV}) = 0.12$ and
$N_f$ = 5.

The parametrization of the running coupling $\asq$ with $\Lambda$ instead of 
$\as (\mu^2)$ has become a common standard, see e.g. reference~\cite{pdg}, and will
also be adapted in this review. 
While being a convenient and well-used choice, however, this parametrization has
several shortcomings which the user should be aware of.

First, requiring that $\asq$ must be continuous when crossing
a quark threshold\footnote{Strictly speaking, {\em physical observables} $\R$ 
rather than $\as$ must be continuous, which may lead to small discontinuities
in $\asq$ at quark thresholds in finite order perturbation theory; see
section~2.5.},
$\Lambda$ actually depends on the number of active quark flavours. 
Secondly, $\Lambda$ depends on the renormalization scheme, see e.g. reference
\cite{collins-book}. 
In this review, the so-called \oq modified minimal
subtraction scheme" ($\msbar$)
\cite{msbar} will be adopted, which also has become a common standard~\cite{pdg}.
$\Lambda$ will therefore be labelled $\lamsb^{(N_f)}$ to indicate these
peculiarities.

In complete 4-loop approximation and using the $\Lambda$-parametrization, the
running coupling is thus given \cite{alphas-4loop} by
\begin{eqnarray} \label{eq-as4loop}
\as (Q^2) &=& \frac{1}{\beta_0  L} 
               - \frac{1}{\beta_0^3 L^2} \beta_1 \ln  L  \nonumber \\
          &+& \frac{1}{\beta_0^3 L^3} \left( \frac{\beta_1^2}{\beta_0^2}
              \left( \ln^2 L - \ln L - 1 \right) + \frac{\beta_2}{\beta_0}
               \right) \nonumber \\
          &+& \frac{1}{\beta_0^4 L^4} \left( \frac{\beta_1^3}{\beta_0^3}
              \left( - \ln^3 L + \frac{5}{2} \ln^2 L + 2 \ln L - \frac{1}{2}
              \right) - 3 \frac{\beta_1 \beta_2}{\beta_0^2} \ln L
              + \frac{\beta_3}{2 \beta_0} \right)  \nonumber \\
\  \ 
\end{eqnarray}
\noindent
where $L = Q^ 2 / \lamsb^2 $.
The first line of equation~\ref{eq-as4loop} includes the 1- and the 2-loop
coefficients, the second line is the 3-loop and the third line is the 4-loop
correction, respectively.

\begin{figure}[ht]
\begin{center}
\epsfxsize12.0cm\epsffile{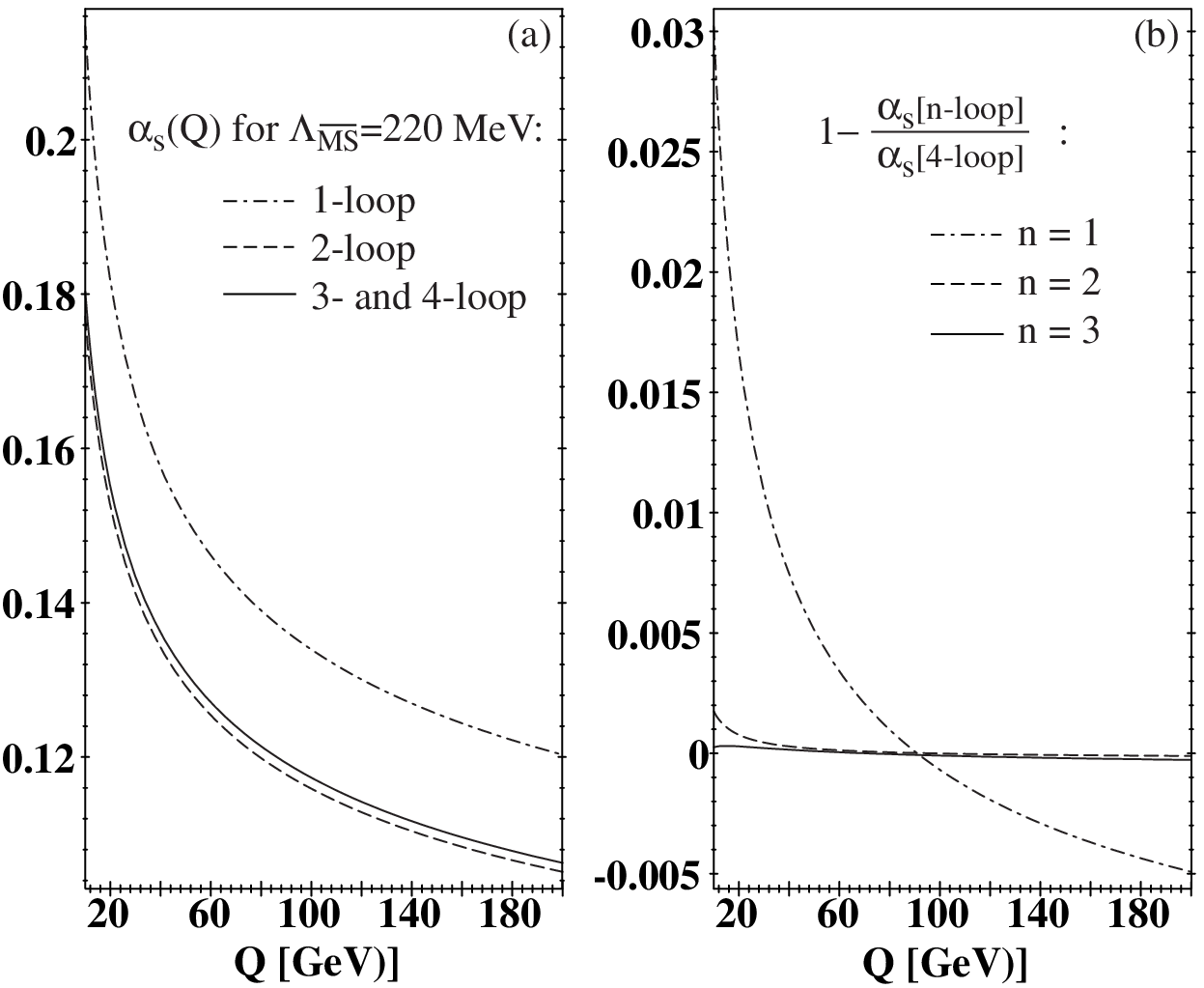} 
\end{center}
\caption{(a) The running of $\as (Q)$, according to
equation~\protect\ref{eq-as4loop},  in 1-,
2- and 3-loop approximation, for $N_f = 5$ and with an identical value of 
$\lamsb = 0.22$~GeV.
The 4-loop prediction is indistinguishable from the 3-loop curve. 
(b) Fractional difference between the 4-loop and the 1-, 2- and 3-loop
presentations of
$\as (Q)$, for $N_f = 5$ and $\lamsb$ chosen such that, in each order, $\amz
= 0.119$.
\label{fig:asq-orders}}
\end{figure}

The functional form of $\as (Q)$, for the
1-, the 2- and the 3-loop approximation of equation~\ref{eq-as4loop}, each
with
$\Lambda = 0.220$~GeV, is shown in figure~\ref{fig:asq-orders}(a).
As can be seen, there is an almost 15\% decrease of $\as$ when changing from 1-loop
to 2-loop approximation, for the same value of $\Lambda$.
The difference
between the 2-loop and the 3-loop prediction is only about 1-2\%, and
is less than 0.01\% between the 3-loop and the 4-loop presentation which cannot
be resolved in the figure.

The fractional difference in the energy dependence of $\as$,
$\frac{(\as^{(4-loop)} - \as^{(n-loop)})}{\as^{(4-loop)}}$, for $n$~=~1,~2 and~3,
is presented in figure~\ref{fig:asq-orders}(b).
Here, in contrast to figure~\ref{fig:asq-orders}(a), the values of $\lamsb$ were
chosen such that $\amz = 0.119$ in each order,
i.e., $\lamsb = 93$~MeV (1-loop), $\lamsb = 239$~MeV (2-loop), and
$\lamsb = 220$~MeV (3- and 4-loop).
Only the 1-loop approximation shows sizeable differences of up to
several per cent, in the energy and parameter range chosen, while the 2- and
3-loop approximation already reproduce the energy dependence of the best, i.e.
4-loop, prediction quite accurately.

\subsection{Quark masses and thresholds}

So far in this discussion, finite quark masses $m_q$ were neglected, assuming that
both the physical and the renormalization scales $Q^2$ and $\mu^2$, respectively,
are larger than any other relevant energy or mass scale involved in the problem.
This is, however, not entirely correct, since there are several
QCD studies and $\as$ determinations at energy scales around
the charm- and bottom-quark masses of about 1.5 and 4.7~GeV, respectively.

Finite quark masses may have two major effects on actual QCD studies:
Firstly, quark masses will alter the perturbative predictions of
observables $\cal{R}$.
While phase space effects which are introduced by massive quarks can  often
be studied using hadronization models and Monte Carlo simulation techniques,
explicit quark mass corrections in higher than leading perturbative order are
available only for  
jet production \cite{qmasses} and for total hadronic cross sections
\cite{xmasses,xmasses2} in
$\epem$ annihilation.

Secondly, any quark-mass dependence of $\cal{R}$ will add another term
$\mu^2 \frac{\partial m}{\partial \mu^2}\frac{\partial}{\partial m} {\cal R}$
to equation~\ref{eq-muindependence}, which leads to energy-dependent, running quark
masses, $m_q (Q^2)$, in a similar way as the running coupling $\as (Q^2)$ was
obtained, see e.g. reference~\cite{ellis-book}.

In addition to these effects, $\as$ indirectly also depends on the quark masses,
through the dependence of the $\beta$ coefficients on the effective number of
quarks flavours, $N_f$, with $m_q \ll \mu$.
Constructing an effective theory for, say, ($N_f$-1) quark flavours which must be
consistent with the $N_f$ quark flavours theory at the heavy quark threshold
$\mu^{(N_f)} \sim {\cal O} (m_q)$, results in matching conditions for the $\as$
values of the ($N_f$-1)- and the $N_f$-quark flavours theories \cite{bernreuther}.

In leading and in next-to-leading order, the matching condition is
$\as^{(N_f-1)} = \as^{N_f}$.
In higher orders and the $\overline{MS}$ scheme, however, nontrivial matching
conditions apply \cite{bernreuther,larin,alphas-4loop}.
Formally these are, if the energy evolution of $\as$ is performed in $n^{th}$
order, of order ($n-1$).

The matching scale $\mu^{(N_f)}$ can be chosen in terms of the (running)
$\overline{MS}$ mass $ m_q ( \mu_q )$, or of
the constant, so-called pole mass $ M_q $.
For both cases, the relevant matching conditions in NNLO are given in
\cite{alphas-4loop}. 
These expressions have a particluarly simple form for the 
choice\footnote{The results of reference~\cite{alphas-4loop} are also valid for
other relations between $\mu^{(N_f)}$ and $m_q$ or $M_q$, as e.g. $\mu^{(N_f)}
= 2 M_q$. 
For 3-loop matching, however, practical differences due to the freedom of this
choice are negligible.}
$\mu^{(N_f)} = m_q (m_q)$ or $\mu^{(N_f)} = M_q$.
In this report, the latter choice will be used to perform 3-loop matching
at the heavy quark pole masses, in which case the matching condition reads, with
$a = \as^{(N_f)} / \pi$ and $a' = \as^{(N_f-1)} / \pi$:

\begin{equation} \label{Mq-matching}
\frac{a'}{a} = 1 + C_2\ a^2 + C_3\ a^3 \ ,
\end{equation}
\noindent
where $C_2 = - 0.291667$ and $C_3 = -5.32389 + (N_f-1)\cdot 0.26247$
\cite{alphas-4loop}.

\begin{figure}[ht]
\begin{center}
\epsfxsize12.0cm\epsffile{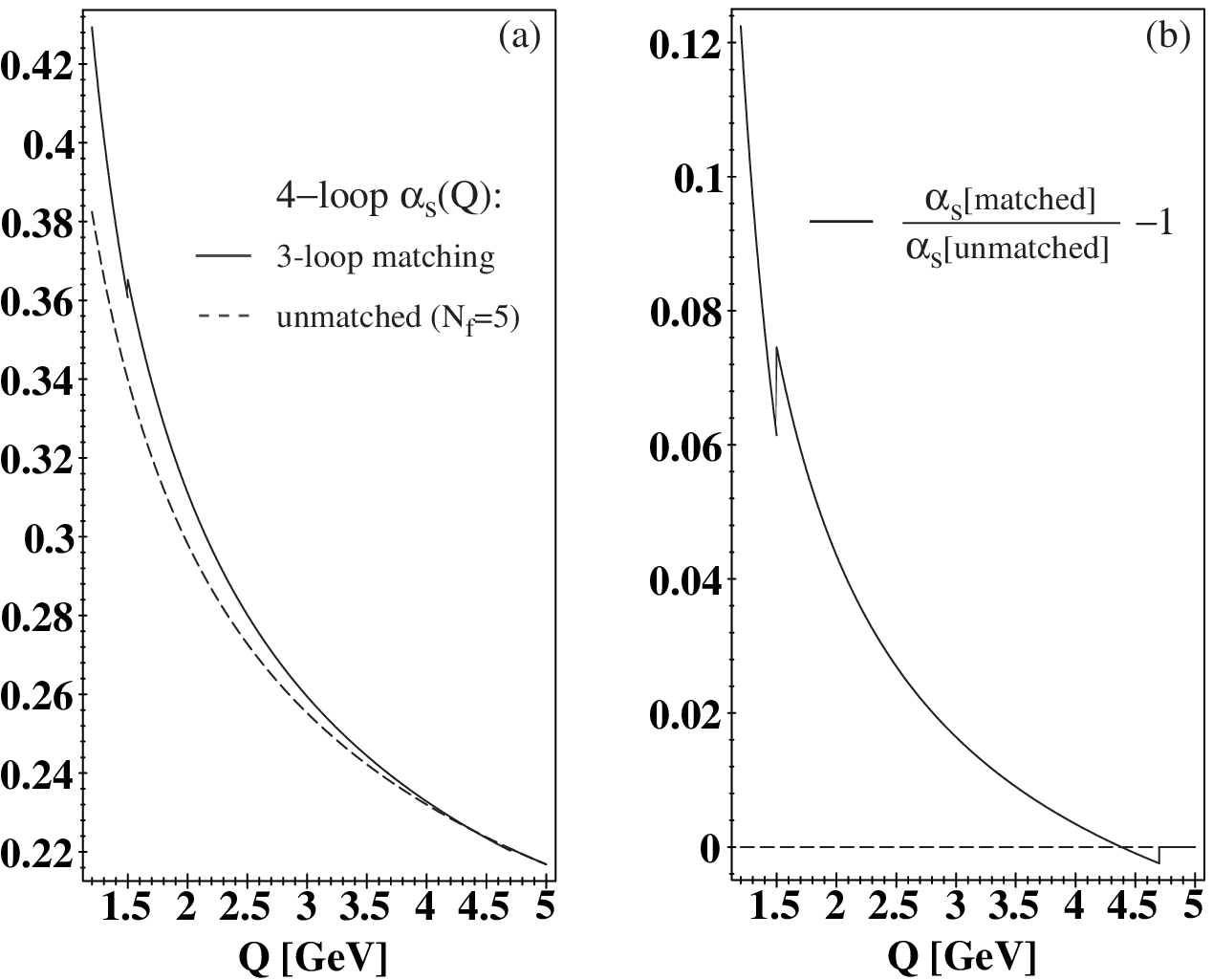} 
\end{center}
\caption{(a) 4-loop running of $\as (Q)$ with 3-loop quark threshold matching
according to equations~\protect\ref{eq-as4loop} and~\protect\ref{Mq-matching}, with
$\lamsb^{(N_f = 5)}$~=~220~MeV and charm- and bottom-quark thresholds at the
pole masses,
$\mu^{(N_f = 4)}_c \equiv M_c = 1.5$~GeV and $\mu^{(N_f = 5)}_b \equiv M_b =
4.7$~GeV (full line),
compared with the unmatched 4-loop result (dashed line).
(b) The fractional difference between the two curves in (a).
\label{fig:as-match}}
\end{figure}

The 4-loop prediction for the running $\as$, using equation~\ref{eq-as4loop} with
$\lamsb^{(N_f = 5)}$~=~220~MeV and 3-loop matching at the charm- and bottom-quark
pole masses,
$\mu^{(N_f = 4)}_c = M_c = 1.5$~GeV and $\mu^{(N_f = 5)}_b = M_b = 4.7$~GeV,
is illustrated in figure~\ref{fig:as-match}a (full line).
Small discontinuities at the quark thresholds can be seen, such that
$\as^{(N_f-1)} < \as^{(N_f)}$ by about 2~per mille at the bottom- and about
1~per cent at the charm-quark threshold.
The corresponding values of $\lamsb$ are $\lamsb^{(N_f=4)} = 305$~MeV and
$\lamsb^{(N_f=3)} = 346$~MeV \cite{as4}.
Comparison with the 4-loop prediction, without applying
threshold matching and for $\lamsb^{(N_f = 5)}$~=~220~MeV and $N_f$~=~5 
throughout (dashed
line) demonstrates that, in spite of the discontinuities, the matched calculation
shows a steeper rise towards smaller energies because of the larger values of
$\lamsb^{(N_f=4)}$ and $\lamsb^{(N_f=3)}$.

The size of discontinuities and the changes of slopes are more clearly demonstrated
in figure~\ref{fig:as-match}b, where the fractional difference between the two
curves from figure~\ref{fig:as-match}a, i.e. between the matched and the unmatched
calculation, is presented.
Note that the step function of $\as$ is not an effect
which can be measured --- the steps are artifacts of the truncated
perturbation theory and the requirement that predictions for observables at
energy scales around the matching point must be consistent and independent of the
two possible choices of (neighbouring) values of $N_f$.

\subsection{Perturbative predictions of physical quantities}

In practice, $\as$ is not an `observable' by itself.
Values of $\as (\mu^2)$ are
determined from measurements of observables $\cal{R}$ for which perturbative
predictions exist. 
These are usually given by a power series in $\as (\mu^2 )$, like
\begin{eqnarray} \label{eq-rseries}
{\cal R}(Q^2) &=&   P_{l} \sum_{n} R_n \as^n \nonumber \\
              &=& P_l \left( R_0 + R_1 \as (\mu^2) + R_2 (Q^2 / \mu^2 ) \as^2
(\mu^2 ) + ...\right) \  ,
\end{eqnarray}
\noindent 
where $R_n$ are the $n_{th}$ order coefficients of the perturbation series and
$P_l R_0$ denotes the lowest-order value of $\cal R$.

For processes which involve gluons already in lowest order perturbation theory,
$P_l$ itself may include (powers of) $\as$. 
For instance, this happens in case of the hadronic decay
width of heavy Quarkonia, $\Gamma (\Upsilon \rightarrow ggg \rightarrow\
hadrons)$ for which $P_l \propto \as^3$. 
If no gluons are involved in lowest order, as e.g. in $\epem \rightarrow\
q\overline{q} \rightarrow \ hadrons$ or in deep inelastic scattering processes,
$P_l R_0$ is a constant and the usual choice is $P_l \equiv 1$.
$R_0$ is called the {\it lowest order} coefficient and $R_1$ is the
{\it leading order} (LO) coefficient.
Following this naming convention, $R_2$ is the {\it next-to-leading order}
(NLO) and $R_3$ is the {\it next-to-next-to-leading order}
(NNLO) coefficient.

QCD calculations in NLO perturbation theory are
available for many observables $\R$ in high energy particle reactions;
calculations including the complete NNLO are
available for some totally inclusive quantities,
like the total hadronic cross section in $\epem \rightarrow\ hadrons$, moments
and sum rules of structure functions in deep inelastic scattering processes and
the hadronic decay width of the $\tau$ lepton.
The complicated nature of QCD, due to the process of gluon self-coupling and the
resulting large number of Feynman diagrams in higher orders of perturbation
theory, so far limited the number of QCD calculations in complete NNLO.

An alternative approach to calculating higher order corrections is
based on the
resummation of leading logarithms which arise from soft and collinear
singularities in gluon emission \cite{resummation}.   
In such a case, the effective expansion parameter is 
$\as L^2$ rather than $\as$, 
where $L=\ln(1/\cal{R})$ and $\cal{R}$ is some generic observable which
tends to zero in lowest order.  
For small values of $\cal{R}$,  $\as L^2$
becomes large, and therefore these terms should be 
known to all orders in $\as$ 
if a reliable prediction of $\cal{R}$ is to be obtained.  
For certain observables
it has proved possible to sum up both the leading and next-to-leading
logarithms, which is referred to as the `Next-to-Leading Log
Approximation' or NLLA.   

For observables $\cal{R}$ 
for which resummation is possible, 
the $cumulative$ cross-section $\Sigma (\cal{R})$ may be written as
\begin{equation}
\Sigma ({\cal R}) \equiv \int_0^{\cal R}
\frac{1}{\sigma}\frac{d\sigma}{d{\cal R}} d{\cal R} = C(\as) \exp \left[ G(\as,L)
\right] + D(\as,{\cal R}) \ ,
\end{equation}
where $D(\as,\cal{R})$ is a remainder function which should 
vanish as ${\cal R} \rightarrow 0$, and
\begin{eqnarray}
C(\as) &=& 1 + \sum_{n=1}^{\infty} C_n \asb^n  \\ \nonumber
G(\as,L) &=& \sum_{n=1}^{\infty}\sum_{m=1}^{n+1}G_{nm}\asb^nL^m \\ \nonumber
&\equiv&
L g_1(\as L) + g_2(\as L) + \as g_3(\as L) + \as^2 g_4(\as L) \cdots \ ,
\end{eqnarray}
with $\asb \equiv (\as/2\pi)$.
The functions $L g_1$ and $g_2$ represent the sums of 
the leading and next-to-leading logarithms respectively, to all 
orders in $\as$.

Resummation of leading and next-to-leading logarithms provides predictions
which include leading terms of all perturbative orders; however, they are not
complete in the sense of fixed-order predictions because the latter also
include sub-leading logarithms and non-logarithmic terms.
This is illustrated in Table~\ref{tab:resum},
where the NLLA calculations
provide  the sum of all terms of the first two columns, while NLO ($\oaa$)
calculations yield the sums of the terms in the first two rows.  

\begin{table}[htb]
\caption{\label{tab:resum} Decomposition of the cumulative cross-section, $\ln
\Sigma({\cal R})$,  in powers of $\asb=(\as/2\pi)$ and $L=\ln(1/{\cal R})$. }
\begin{center}
\renewcommand{\arraystretch}{1.5}
\begin{tabular} {r|c|c|c|c|l} \cline{2-5}
   &  Leading & Next-to- & Subleading & Non-log. & \\
   &   logs   &     Leading logs        &   logs     & terms & \\
\cline{2-5}
 $\ln \Sigma (\cal{R})$ = & $G_{12}\asb  L^2$ &$+\, G_{11}\asb  L $ & 
    & $+\,\as\cal O $(1) 
& $\oa $ \\
           & $+\,G_{23}\asb^2L^3 $ &$+\,G_{22}\asb^2L^2$ &
                          $+\,G_{21}\asb^2L$ &$+\,\as^2\cal O$(1)   
& $\oaa $ \\ 
           & $+\,G_{34}\asb^3L^4$ & $+\,G_{33}\asb^3L^3$ &
               $+\,G_{32}\asb^3L^2+\cdots$ & $ +\cdots$ & $\oaaa$\\
           &  $+\cdots$  & $+\cdots$ & $+\cdots$  & $+\cdots$  
           &  \ \ \vdots \\ 
\cline{2-5}
            = & $L g_1(\as L) $ & $ +\,g_2(\as L) $ & $ +\cdots$ & $+\cdots$ \\
\cline{2-5}
\end{tabular}   
\end{center}
\end{table}

If the maximum available information from both resummed and fixed order
calculations shall be utilized, they must be added such
that double counting of terms which are common to both
these calculations is avoided.
This can be achieved involving several (approximate)
`matching schemes'.
For instance, terms to $\oaa$ in the NLLA
expression for $\ln \Sigma ({\cal R})$ (c.f. table~\ref{tab:resum})
can be removed and replaced by the full
expression in exact $\oaa$.
This procedure is called `$\ln R$ matching'; it results in `resummed $\oaa$'
predictions which should be superior to the fixed $\oaa$ case because more
higher order correction terms are included. 

The perturbative order up to which QCD predictions for different
processes and observables are available, is indicated in 
Table~\ref{tab:astab}, at the end of this review.

\subsection{Renormalization scale dependence - infinite discomfort
in finite order.}

The principal independence of a physical observable $\R$
from the choice of the renormalization scale $\mu$ was expressed in
equation~\ref{eq-muindependence}.
Replacing $\as$ by $\as (\mu^2)$, using
equation~\ref{eq-rge}, and inserting the perturbative expansion of $\R$ 
(equation~\ref{eq-rseries}) into equation~\ref{eq-muindependence} results in
\begin{eqnarray} \label{eq-muindependence2}
0 = \mu^2 \frac{\partial R_0}{\partial \mu^2} 
    + \as (\mu^2) \mu^2 \frac{\partial R_1}{\partial \mu^2} \nonumber 
    &+& \as^2 (\mu^2) \left[ \mu^2 \frac{\partial R_2}{\partial \mu^2} -
    R_1 \beta_0 \right]  \nonumber \\
    &+& \as^3 (\mu^2) \left[ \mu^2 \frac{\partial R_3}{\partial \mu^2} -
    [R_1 \beta_1 + 2 R_2 \beta_0] \right] \nonumber \\
    &+& {\cal O} (\as^4) \ .
\end{eqnarray}
Solving this relation requires that the coefficients of $\as^n (\mu^2)$ vanish
for each order $n$.
With an appropriate choice of integration limits one thus obtains
\begin{eqnarray} \label{eq-R-mudependence}
&R_0& = {\rm const.}\ , \nonumber \\
&R_1& = {\rm const.}\ , \nonumber \\
&R_2& \left(\frac{Q^2}{\mu^2}\right) = R_2 (1) - \beta_0 R_1 \ln
\frac{Q^2}{\mu^2}\ , \nonumber \\
&R_3& \left(\frac{Q^2}{\mu^2}\right) = R_3 (1) - [ 2 R_2(1) \beta_0 + R_1 \beta_1
]  \ln \frac{Q^2}{\mu^2} + R_1\beta_0^2 \ln^2 \frac{Q^2}{\mu^2}
\end{eqnarray}
\noindent
as a solution of equation~\ref{eq-muindependence2}.

In other words, invariance of the complete perturbation series
against the choice of the renormalization scale $\mu^2$ implies that the
coefficients
$R_n$, except $R_0$ and $R_1$, explicitly depend on $\mu^2$.
In infinite order, the renormalization scale dependence of $\as$ and of the
coefficients $R_n$ cancel; in any finite (truncated) order, however, the
cancellation is not perfect, such that all realistic perturbative QCD
predictions include an explicit dependence on the choice of the
renormalization scale.

This renormalization scale dependence is most pronounced in leading order QCD
because $R_1$ does not depend on $\mu$ and thus, there is no cancellation of
the (logarithmic) scale dependence of $\as (\mu^2)$ at all.
Only in next-to-leading and higher orders, the scale dependence of the
coefficients $R_n$, for $n \ge 2$, partly cancels that of $\as (\mu^2)$.
In general, the degree of cancellation improves with the inclusion of higher
orders in the perturbation series of $\R$.

\begin{figure}[ht]
\begin{center}
\epsfxsize13.5cm\epsffile{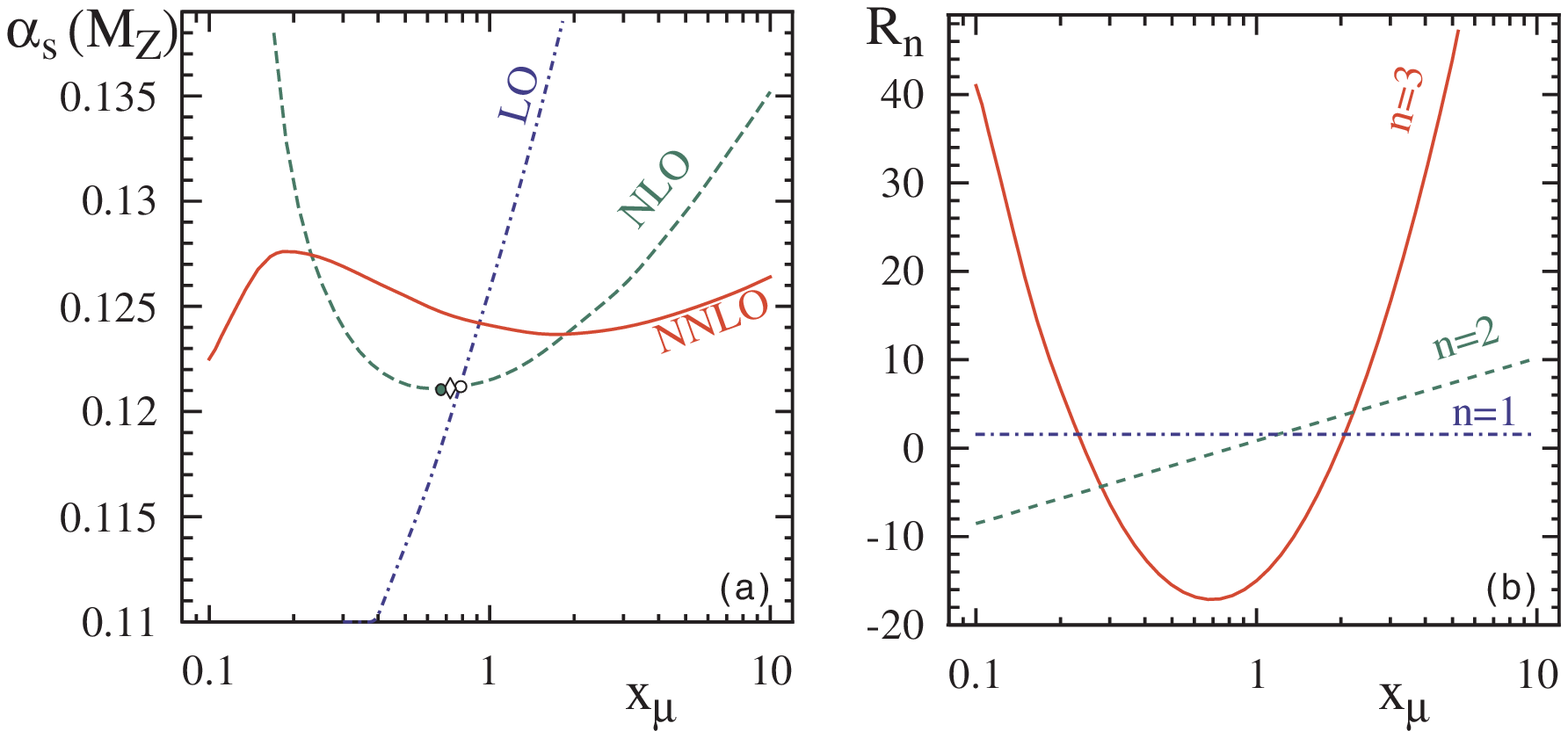} 
\end{center}
\caption{(a) $\amz$ determined from the scaled hadronic
width of the $\z0$, $\rz = 20.768$, in leading, next-to-leading and in
next-to-next-to leading order QCD, as a function of the renormalization scale
factor $x_{\mu} = \mu / \mz$.
NLO solutions according to the PMS ($\bullet$), the EC ($\circ$) and the BLM
($\diamond$) scale optimization methods are
marked.
(b) Scale dependence of the 
QCD coefficients $R_n$ of $\rz$.
\label{fig:as-Rz}}
\end{figure}

A practical example of the scale dependence of $\as$, determined from the
measured value of the scaled hadronic decay width of the $\z0$ boson
\cite{lep-ew},
\begin{equation} \label{eq-Rz-def}
\rz = \frac{\Gamma (\z0 \rightarrow {\rm hadrons} )}{\Gamma (\z0 \rightarrow
{\rm leptons} )} = 20.768 \pm 0.0024\ ,
\end{equation}
is shown in figure~\ref{fig:as-Rz}.
QCD predictions of $\rz$ are available in complete NNLO
\cite{rznnlo}.
Including further corrections like quark mass and
non-factorizable electroweak and QCD effects, these predictions can be
parametrised \cite{r-param} to
\begin{equation} \label{eq-Rz-coeff}
\rz = 19.934  \left[ 1 + 1.045 \frac{\as (\mu)}{\pi} +  0.94
\left[\frac{\as (\mu)}{\pi}\right]^2
 - 15\left[\frac{\as (\mu)}{\pi}\right]^3 \right]\ ,
\end{equation}
for $\mu \equiv \mz$; see section~4.3 for more details.
In order to demonstrate the scale dependence of $\as$,
$\as (\mu)$ is determined from 
equations~\ref{eq-Rz-def} and~\ref{eq-Rz-coeff} in LO, in NLO
and in NNLO, as a function of the scaled renormalization scale, $x_{\mu} = \mu /
\mz$, using the appropriate, scale dependent QCD coefficients $R_n$
according to equation~\ref{eq-R-mudependence}.
$\amz$ is then calculated from $\as (\mu )$ using equation~\ref{eq-as4loop} in
1-loop expansion for the LO, in 2-loop for the NLO and in 3-loop for the NNLO
case.

The resulting scale dependence of $\amz$ is displayed in
figure~\ref{fig:as-Rz}(a);
the scale dependence of the QCD coefficients $R_1$, $R_2$ and
$R_3$ is given in figure~\ref{fig:as-Rz}(b).
The scale dependence of $\as$ is visibly reduced if
higher orders are included:
limiting renormalization scales to factors from 1/5 to 5 of 
the \oq physical" scale  $Q \equiv \mz$,
results in changes of $\amz$ of about 40\% in LO, 8\% in NLO and 2.5\% in NNLO.
Beyond these limits, however, the scale dependence even in NNLO QCD diverges,
such that it seems not to be meaningful to consider yet larger ranges of
scale factors $\xmu$ - although there is no rule, from first principles, which
limits scale variations to a definite range.

There are several proposals to optimize or to fix the
renormalization scale in finite order perturbation
theory.
An intuitive approach is to apply the general scale insensitivity
requirement, equation~\ref{eq-muindependence}, which is strictly valid in infinite
order, to calculations in any finite (beyond leading) order.
This is Stevenson's principle of minimal sensitivity (PMS)\cite{stevenson},
where, for each observable $\R$, the optimal scale
is defined by d$\R/\rm{d}\mu^2 = 0$.
The PMS solution for $\rz$ in NLO is marked in figure~\ref{fig:as-Rz}(a)  [filled
circle].

In the \oq effective charge" method (EC) of Grunberg
\cite{grunberg}, all higher (i.e. beyond leading) order coefficients $R_n$
vanish.
The EC solution in NLO is also marked in figure~\ref{fig:as-Rz}(a) [open
circle].

Brodsky, Lepage and
Mackenzie (BLM) \cite{blm} propose to choose $\mu$
such that the NLO coefficient $R_2$ is independent of $N_f$, the number of
active quark flavours. 
The BLM solution is marked in figure~\ref{fig:as-Rz}(a) [open rhombus].

PMS, EC and BLM all provide unambiguous solutions in NLO; in the case of
$\rz$, they are very close to each other but the NNLO prediction does not
exactly match these results: the value of $\as$ from \oq optimized" NLO can only
be reached in NNLO at very small values of $x_{\mu}$.

While in NLO QCD it is sufficient to vary the renormalization scale in order to
assess the full spectrum of variation offered by the renormalization procedure,
in NNLO $both$ scale- and scheme-dependences must be taken into account.
PMS, EC and BLM can then be generalized to NNLO and beyond, however do
not lead to unambiguous results anymore.
The whole issue of theoretical scale optimizations was and still is vividly
discussed in the literature, see e.g. \cite{scale-various}.

Another approach is motivated by an experimental point of view:
ideally, the optimal choice of $\mu$ in a given order of perturbation theory
should reproduce or, at least, closely approach the unknown all-order result.
Measurements of observables should inherently include all orders of perturbation
theory; therefore it may be possible to extract the optimal choice of $\mu$
in a given, finite order calculation from fits to the experimental data.
Indeed, experimental scale optimization is possible in
cases where $\as$ is determined from differential $distributions$ of observables,
and often leads to a remarkable consistency between results obtained from different
observables \cite{siggiscale,d-old-as,o-grand-as,d-as}.

So far, none of the methods described above was generally accepted as $the$ 
preferred method to optimize or fix renormalistaion scales in finite order
calculations.
Instead, in experimental determinations of $\as$, systematic uncertainties due
to unknown higher order contributions are usually defined by allowing the
renormalization scale $\mu$ to vary in \oq reasonable" ranges.

Unfortunately, there is no common agreement on what should be a reasonable
range, and the significance and interpretation of such procedures is unclear.
Furthermore, as indicated above, scale changes alone are not sufficient to
investigate higher order uncertainties in NNLO predictions, where renormalization
scheme dependences should also be accounted for.

\subsection{Lattice QCD}

At large distances or low momentum transfers, $\as$ becomes large and
application of perturbation theory becomes inappropriate.
In this regime, nonperturbative methods must be used to describe processes
of the Strong Interaction.
Lattice QCD is one of the most developed nonperturbative methods which is used
to calculate, for instance, hadron masses, hadron mass splittings and QCD matrix
elements.
In Lattice QCD, field operators are applied on a discrete, 4-dimensional
Euclidean space-time of hypercubes with side length $a$.

Energy levels of heavy quarkonia systems ($Q\overline{Q}$) 
calculated using lattice QCD depend on the heavy quark mass $M_Q$ and on
$\as$.
Comparison with measurements of quarkonia mass splittings then allows to
determine $\as$, which at this point is not given in the $\msbar$
renormalization scheme but is based on other suitable definitions of $\as$ on
the lattice. 
Conversion from the lattice to the $\msbar$ coupling at high energies can be
done using an expansion in second or in third order perturbation theory.
A review of methods to determine $\as$ from lattice investigations is given in
\cite{weisz}

So far, calculations exist which either neglect light quark loop contributions
($N_f$=0; \oq quenched approximation") or which include two
light quark flavours ($N_f$=2); the latter allow
extrapolation to $N_f$=3.
Uncertainties in this extrapolation, the limited order to which the conversion to
the $\msbar$ scheme is known, limited Monte-Carlo statistics
and corrections for light quark masses lead to theoretical uncertainties
of final ($\msbar$) values of $\amz$.

\section{Results from deep inelasting scattering processes} \label{sec:dis}

Measurements of the violation
of Bjorken scaling in deep inelastic lepton-nucleon scattering belong
to the earliest methods to determine $\as$.
The first {\em significant} determinations of $\as$, i.e. 
those based at least on next-to-leading order (NLO)
perturbative QCD prediction, started to emerge in 1979 \cite{yndurain-79}.

Today, a large number of results is available, from data in the energy ($Q^2$)
range of a few to several thousand GeV$^2$, using electron-, muon- and
neutrino-beams on various fixed target materials (H, D, C, Fe and CaCO$_3$), as
well as electron-proton or positron-proton colliding beams at HERA. 
In addition to scaling violations of structure functions, $\as$ is also
determined from moments of structure functions, from QCD sum rules and --- as in
$\epem$ annihilation --- from hadronic jet production and event shapes.

\subsection{Scaling violations and moments of structure functions}

Cross sections of physical processes in lepton-nucleon scattering and
in hadron-hadron collisions depend on the quark- and
gluon-densities in the nucleon.
Assuming factorization between short-distance, hard scattering processes which
can be calculated using QCD perturbation theory, and low-energy or long-range
processes which are not accessible by perturbative methods,
such cross sections are 
parametrized by a set of structure functions $F_i$ ($i$= 1,2,3).
The transition between the long- and the short-range regimes is defined by an
arbitrary factorization scale $\mu_f$, which --- in general --- is independent
from the renormalization scale $\mu$, but has similar features as the latter:
the higher order coefficients of the perturbative QCD series for physical cross
sections depend on $\mu_f$ in such a way that the cross section to all orders
must be independent of $\mu_f$, i.e. $\partial \sigma / \partial \mu_f = 0$.
To simplify application of theory to experimental measurements, the assumption
$\mu_f = \mu$ is usually made, with $\mu \equiv Q$ as the standard choice of
scales.

In the naive quark-parton model, i.e. neglecting gluons and QCD,
the differential cross sections for electromagnetic charged lepton
(electron or muon) or weak neutrino-proton scattering 
off an unpolarized proton target are written
\begin{eqnarray} \label{eq-dis}
\frac{{\rm d}^2 \sigma^{em}}{{\rm d}x {\rm d}y} &\sim& \frac{1 + (1-y)^2}{2} 2x F_1
+ (1-y)(F^{em}_2 - 2xF^{em}_1) - \frac{M}{2E} xyF^{em}_2\ ; \nonumber \\
\frac{{\rm d}^2 \sigma^{\nu}}{{\rm d}x {\rm d}y} &\sim& \left( 1-y-\frac{M}{2E} xy
\right) F^{\nu}_2 + y^2 xF^{\nu}_1 + y\left( 1-\frac{1}{2} y\right) xF^{\nu}_3\ ,
\end{eqnarray}
where $x = \frac{Q^2}{2M(E-E')}$
is the momentum fraction of the nucleon
carried by the struck parton,
$y = 1 - E'/E$,
$Q^2$ is the negative quadratic momentum transfer in the scattering process, and
$M$, $E$ and $E'$ are the mass of the proton and the lepton energies before and
after the scattering, respectively, in the rest frame of the proton.
In the quark-parton model, these structure functions consist of combinations
of the quark- and antiquark densities $q(x)$ and $\overline{q}(x)$ for both
valence- (u,d) and sea-quarks (s,c).

In QCD, the gluon content of the proton as well as higher order diagrams describing
photon/gluon scattering, $\gamma g \rightarrow q \overline{q}$, and gluon radiation
off quarks must be taken into account.
Quark- and gluon-densities, the structure functions $F_i$ and physical
cross sections become energy ($Q^2$) dependent.
QCD thus predicts, departing from the naive quark-parton model,
scaling violations in physical cross sections, which are associated
with the radiation of gluons.
While perturbative QCD cannot predict the functional form of parton densities and
structure functions, their energy evolution is described by the
so-called DGLAP equations \cite{dglap,ap}.
 
Structure functions contain,
apart from terms whose energy dependence is given by perturbative QCD, which so
far is known up to complete NLO \cite{sf-nlo},
so-called \oq higher twist" contributions (HT).
The leading higher twist terms are proportional to $1/Q^2$; they are
numerically important at low $Q^2 < \cal{O}$(few~GeV$^2)$ and at very large $x
\simeq 1$.

Historically,
precise results of $\as$, from the logarithmic slopes of 
$F_2 (x,Q^2)$ and in NLO QCD, were obtained 
from a combined analysis~\cite{f2-pion} of SLAC
and BCDMS data, in a $Q^2$ range from 0.5 to 260~GeV$^2$, giving $\amz = 0.113 \pm
0.005$. 
The error includes theoretical (scale) uncertainties of $\pm 0.004$
\cite{dis-scale}. 
Higher twist terms as well as the gluon distribution were simultaneously 
determined in this analysis. 

The CCFR colaboration obtained $\as$ from $\nu$-nucleon scattering and
a fit to the non-singlet structure
function $F_3 (x,Q^2)$~\cite{xf3-pion}, which is independent of the poorly known gluon
distribution.
In order to increase statistical precision, $F_3$
was substituted by $F_2$ at large $x$ where gluons do not contribute
much, giving $\amz = 0.111 \pm 0.002 {\rm (stat.)} \pm 0.003  {\rm (sys.)}\pm 0.004
{\rm (theo.)}$.

These two results were, for several years, the most significant from deep inelastic
lepton-nucleon scattering.
Because they were numerically smaller than typical values obtained from $\epem$
annihilation, $\amz \sim 0.120$ (see Section~4), speculations about possible
reasons for such differences arose.
These speculations came to a halt, at least partly, when the CCFR collaboration
corrected their previous result --- due
to a new energy calibration of the detector --- to $\amz = 0.119 \pm 0.005$
\cite{ccfr}.
Further results from $F_2$, 
in a study \cite{ball} of the HERA data at small $x$ and $Q^2 < 100~{\rm GeV}^2$,
based on NLO QCD including summations
of all leading and subleading logarithms of $Q^2$ and $1/x$,
lead to $\amz = 0.120 \pm 0.005 \pm
0.009$, where the first error is experimental and the second theoretical.
Finally, a reanalysis of the SLAC, BCDMS and NMC data on $F_2$, taking proper
account of point-to-point correlations, resulted in an increase of $\amz$ from 0.113
to 0.118 \cite{alekhin-f2-1}.
A continuation of this analysis, including recent HERA data and careful studies of
the effects of higher twist terms, was reported to result in 
$\amz = 0.116 \pm 0.003$ \cite{alekhin-f2-2}.

\begin{figure}[ht]
\begin{center}
\epsfxsize12.0cm\epsffile{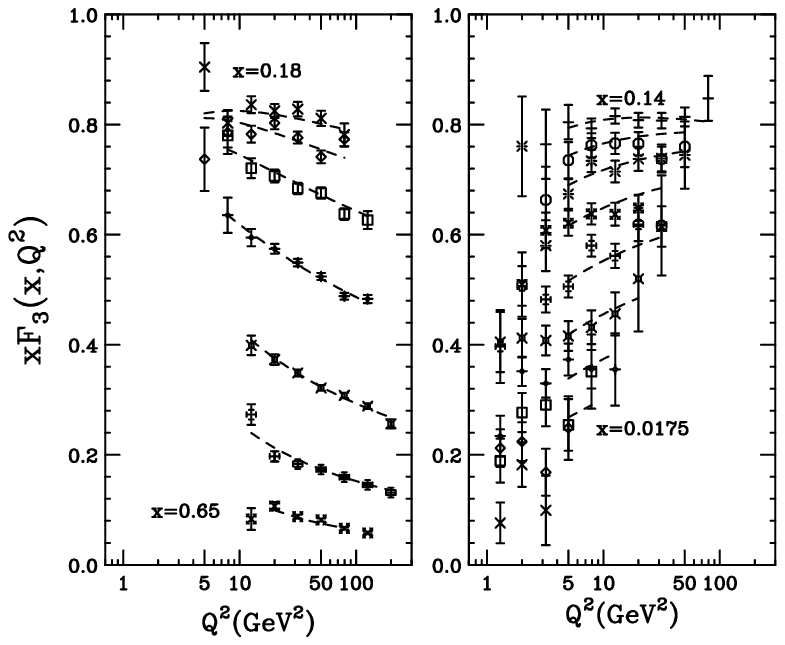} 
\end{center}
\caption{Comparison of CCFR data \protect\cite{ccfr} with NNLO QCD fits
\protect\cite{xf3-recent}.}
\label{fig:xf3}
\end{figure}

Significant progress in this field was achieved with the availability
of NNLO QCD predictions for the non-singlet structure function $F_3$ as well as for
moments of $F_2$.
A new analysis \cite{xf3-recent} of the CCFR data, now in NNLO QCD and from
$xF_3$ alone, resulted in 
$$
\amz = 0.118 \pm 0.002 {\rm (stat.)} \pm 0.005 {\rm
(sys.)}\pm 0.003 {\rm (theo.)}\ ;
$$
see figure~\ref{fig:xf3}. 
This value is taken as the currently
most significant result from structure functions in neutrino-nucleon deep inelastic
scattering, and is included in the final summary, see
Table~\ref{tab:astab}. 
Another recent study was based on Bernstein polynomials and moments of $\F_2$ from
electron- and muon-scattering data, 
including fixed target as well as
HERA colliding beam data in the $Q^2$ range of 2.5 to 230~GeV$^2$.
The result  \cite{f2-recent} was
$$\amz = 0.1172\pm 0.0017 \pm 0.0017\ ,
$$
where the first error is experimental and the second
theoretical. 
The theoretical error includes higher twist effects and an estimate of the NNNLO
corrections.
This result is taken as the final value from $F_2$ in DIS and is added to
the summary in section~7.

From scaling violations of polarized structure functions, based on data from SLAC
and SMC, $\as$ was determined \cite{dispol} in NLO QCD, resulting in
$$\amz = 0.120 ^{+0.004}_{-0.005} {\rm(exp.)} ^{+0.009}_{-0.006}{\rm (theo.)}\ ,$$
which is also added to the final summary.

\subsection{Sum Rules}
 
\begin{figure}[ht]
\begin{center}
\epsfxsize5.0cm\epsffile{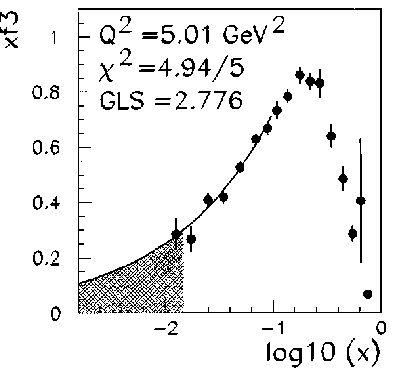} 
\epsfxsize5.4cm\epsffile{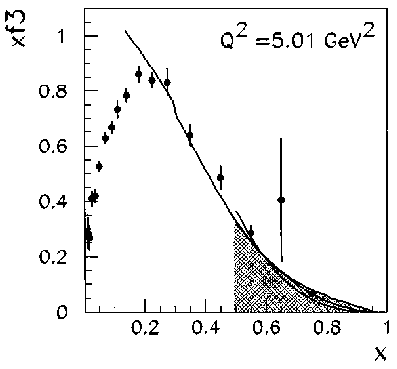}
\end{center}
\caption{$xF_3$ as a function of $x$ at $Q^2 = 5$~GeV$^2$, with a power
law fit to the region $x<0.1$ (left figure; logarithmic $x$-scale) and the shape of
the SLAC $F_2$ data for $x>0.5$ (right figure; linear $x$-scale) which are used to
calculate the GLS integral in the shaded regions (figures from
\cite{gls-recent}).}
\label{fig:gls}
\end{figure}

Perturbative corrections to two inclusive measurements,
namely the Gross-Llewellyn Smith (GLS) sum rule~\cite{gls} for deep inelastic
neutrino scattering and the Bjorken sum rule~\cite{bj-sr} for polarized
structure functions, have been determined to complete
NNLO QCD~\cite{gls-pion,bj-theory}.
The GLS sum rule,
\begin{equation}
{\rm GLS} = \int^1_0 F_3 (x,Q^2) {\rm d}x\ \equiv\ 3 (1 -
\frac{\as}{\pi} + \dots),
\end{equation}
when fitted to data of the CCFR
collaboration~\cite{ccfr-gls} at $Q^2 = 3\ {\rm
GeV}^2$, resulted in~\cite{gls-pion} $\as ( 1.73\ {\rm GeV}) = 0.32 \pm
0.05$.
A recent update of the data and analysis from CCFR \cite{gls-recent},
in the $Q^2$ range from 1 to 15.5~GeV$^2$, gave
$$
\as (1.73\ {\rm GeV}) = 0.28 \pm 0.035 {\rm (stat.)} \pm 0.050
{\rm (sys.)}^{+0.035}_{-0.030}{\rm (theo.)}\ .
$$
The systematic uncertainty
is dominated by the extrapolation of the GLS integral to the regions $x < 0.01$ where
no measurements exist, and to $x>0.5$ which is substituted by  $F_2$ from SLAC
data, see figure~\ref{fig:gls}. 
The theoretical error is dominated by uncertainties in
the higher twist corrections.

The Bjorken polarized sum rule determines that
\begin{equation}
\int_0^1 \left[ g_1^p(x) -  g_1^n(x)\right] {\rm d}x = \frac{1}{3}
\left| \frac{g_A}{g_V}
\right| \left(1 - \frac{\as}{\pi} + \dots \right),
\end{equation}
where the polarized structure functions $g_1(x)$
for protons (superscript $p$)
and neutrons ($n$) are derived from the difference
of cross sections for parallel
and antiparallel polarized targets and ($\mu$ or e) beams;
$g_A$ and $g_V$ are the constants of the neutron weak decay, $g_A / g_V =
-1.26$.
A study~\cite{bjsr} of the CERN SMC~\cite{smc} and the SLAC
E142~\cite{e142} data obtained
$$\as (1.58\ {\rm GeV}) = 0.375^{+0.062}_{-0.081}.$$
No explicit corrections for nonperturbative higher twist where applied
to derive this result, however an estimate of the size of the ${\cal
O}(\as^4)$ terms~\cite{kataev-o4} was taken into account.

\subsection{Jet Production and Event Shapes}
 
Observables parametrizing hadronic event shapes and jet production rates are the
classical inputs for $\as$ studies in $\epem$ annihilation. 
In recent
years, these observables were also studied in high energetic electron- and
positron-proton collisons at HERA, leading to significant determinations of $\as$.

In detail, inclusive as well as differential jet production rates were studied in
the energy range of $Q^2\sim 10$ up to 10000~GeV$^2$ \cite{herajet}, using variants
of jet algorithms from $\epem$ annihilation which are decribed, in more
detail, in section~4. 
In leading order $\as$,
2 + 1 jet events in deep inelastic $e p$ scattering arise from
photon-gluon fusion and from QCD compton processes.
The term `2 + 1 jet' denotes events where two resolved jets can
be identified, in addition to the beam jet from the remnants of the 
incoming proton. 
Previous NLO QCD predictions~\cite{2+1-theory-old} 
which were shown to be imprecise are now replaced by more recent 
calculations~\cite{2+1-theory}.

The results of $\as$ from jet production at HERA \cite{herajet}
can be summarized to 
$$
\amz = 0.118 \pm 0.002 {\rm (stat.)} \pm 0.008{\rm (sys.)} \pm 0.007 {\rm
(theo.)}\ .
$$
The systematic error contains uncertainties from using different jet
algorithms and hadronization models,  and the theoretical error is dominated by
uncertainties from structure functions and from scale variations.

So far, studies of $\as$ from hadronic event shape distributions at HERA
\cite{h1-shapes} are  based on fits to the energy ($Q^2$) evolution of
mean values of shape distributions, using NLO QCD predictions together with
parametrizations of power corrections $1/Q^p$ \cite{dis-power-corr}.
These methods 
do not yet lead to
consistent results of $\as$ from different shape observables \cite{h1-shapes} ---
at least not to the degree of precision which is obtained from \oq classical"
studies of $\as$ from event shape distributions, as e.g. in $\epem$ annihilation;
see Section~4.1 for further discussion. 
Awaiting further progress in theoretical understanding of power
corrections to event shapes, the $\as$ results of these studies will not yet be
included in the final summary of $\as$.

\section{Results from $\epem$ annihilation} \label{sec:e+e-}


In $\epem$ annihilation
reactions, $\as$ is classically determined from hadronic
event shapes, jet production rates and energy correlations for which complete
NLO QCD calculations are available \cite{ert,kn,catani-seymour}.
The first data analysis of this type, from jet rates observed at the PETRA
$\epem$ collider, emerged in 1982 \cite{as-2nd-jade}.
Reviews of early results on $\as$ from $\epem$ annihilations can be found e.g. in
references \cite{slwu,budapest}. 

More recently, this field was extensively covered by the experiments at the LEP
and SLC colliders, see 
e.g.~\cite{sb-catani,altarelli-92,hebbeker-92,annrev,dallas} for reviews of
the first few years of LEP operation.
Using data from LEP and from earlier
$\epem$ colliders at lower c.m. energies, $\as$ can be determined from scaling
violations of fragmentation functions.
Most importantly, precise determinations of
$\as$ are now obtained from the hadronic partial width of $\z0$ decays, from
overall fits to precision electroweak data and from hadronic branching fractions
of $\tau$ lepton decays, which are calculated in complete NNLO QCD.

\subsection{Event shapes, jet rates and energy correlations}

\subsubsection{Observables in (resummed) NLO QCD}

Hadronic event shape variables, jet rates and energy correlations are tools to
study both the amount of gluon radiation and details of the hadronization process.
The definitions of observables which are applied
to hadronic final states of $\epem$ annihilations, like Thrust, Thrust Major
and Minor, Oblateness, jet masses, the jet broadening
measures, energy correlations and jet production rates, are summarized
elsewhere; see e.g.~\cite{kn,bkss,scotland,nijmegen}.
Only two of these, the Thrust observable and the JADE jet algorithm shall be
described here in some more detail:

The Thrust $T$ \cite{thrust} is the normalized sum of the momentum components 
$\vec{p}_i \vec{n}$ of
all particles $i$ of a given event along a specific axis $\vec{n} $; the axis is
chosen such that $T$ is maximized:
$$
T = {\rm max} \left( \frac{\sum_i | \vec{p}_i \vec{n}|}{\sum_i | \vec{p}_i
|}\right) \ .
$$
Thrust assumes values of $T=1$ for perfectly aligned momentum vectors, i.e. for an
ideal event $\epem \rightarrow q\overline{q}$ with two narrow back-to-back jets,
down to $T=0.5$ for a completely spherical distribution of momentum vectors in
the limit of events with many gluons radiated off the initial $q\overline{q}$
pair.

Within the JADE jet algorithm \cite{jadejet},
the scaled pair mass of two resolvable
jets $i$ and $j$,  $y_{ij} = M_{ij}^2 / E_{vis}^2$,
is required to exceed a
threshold value $y_{cut}$, where $E_{vis}$ is the sum of the measured energies of
all particles of an event.
In a recursive process,
the pair of particles or clusters of particles $n$ and $m$
with the smallest value of $y_{nm}$ is replaced  by  (or \oq recombined" into)
a single jet or cluster $k$ with four-momentum $p_k = p_n + p_m$, as long as
$y_{nm} < \yc$.
The procedure is repeated until all pair masses $y_{ij}$
are larger than the jet resolution parameter $\yc$, and the remaining
clusters of particles are called jets.

Several jet recombination schemes and definitions of $M_{ij}$ exist
\cite{kn,bkss,moretti};
the original JADE scheme with
$M_{ij}^2 = 2\cdot E_i\cdot E_j\cdot (1-\cos{\theta_{ij}})$, where $E_i$ and $E_j$
are the energies
of the particles and $\theta_{ij}$ is the angle between them,
and the \oq Durham" scheme \cite{durham,bkss}
with $M_{ij}^2 = 2\cdot {\rm min}( E_i^2, E_j^2)\cdot (1-\cos{\theta_{ij}})$, were
most widely used at LEP, due to their superior features like small sensitivity to
hadronization and particle mass effects \cite{bkss}.

QCD predictions for hadronic event shapes, of jet
production rates and of energy correlations are available in complete NLO
\cite{ert,kn,catani-seymour}.
Differential distributions of such observables $y$ are parametrized as
\begin{equation}
\frac{1}{\sigma_0} \frac{{\rm d}\sigma}{{\rm d} y} = R_1 (y) \as (\mu^2) + 
R_2(y, Q^2/\mu^2) \as^2(\mu^2)\ ,
\end{equation}
where $\sigma_0$ is the total hadronic cross section in leading order.
In addition, for some of the observables, resummation of the leading and
next-to-leading logarithms (NLLA) is available \cite{resummation} which can be
matched with the NLO expressions (resummed NLO); see section~2.6.

In a typical study, measured distributions are corrected for effects of
limited detector acceptance and resolution. 
Fits of QCD predictions to the data are performed after applying hadronization
corrections, as predicted from hadronization models, to the data or to the
theoretical predictions. 
Systematic uncertainties due to these corrections, from
renormalization scale variations and from different matching procedures, are
studied and included in the overall error of the fitted value of $\as$.
No common agreement exists, however, about the range of variations used to
estimate these uncertainties.

Early determinations of $\as$ from event shapes, jet rates and energy
correlations obtained from experiments at the PETRA and PEP colliders were
summarized to $\as = 0.14 \pm 0.02$ at $Q^2 = \ecm^2 \sim (34\ {\rm GeV})^2$
\cite{slwu,budapest}, in NLO QCD.
The results which led to this average did not include
estimates of theoretical uncertainties; however the error
was determined from the scatter of the single results which were based
on different observables, and thus gave a first estimate of
theoretical uncertainties.

\subsubsection{Results in resummed NLO}

A reanalysis of PETRA data, using refined analysis techniques, resummed
NLO QCD calculations, modern
model calculations and including new observables, quite similar to
recent analyses at LEP, resulted in \cite{j-as}
\begin{eqnarray}
\as \left( 22\ {\rm GeV}\right) &=& 0.161\pm 0.009 \pm ^{+0.014}_{-0.006}\ ,
\nonumber \\ 
\as \left( 35\ {\rm GeV}\right) &=& 0.145\pm 0.002 ^{+0.012}_{-0.007}\ ,\ 
{\rm and} \nonumber \\
\as \left( 44\ {\rm GeV}\right) &=& 0.139\pm 0.004 ^{+0.010}_{-0.007},
\nonumber
\end{eqnarray}
where the first errors are statistical and experimental systematic, added in
quadrature, and the second are  theoretical uncertainties. 
These values are taken as the final results in the
PETRA energy range and are included in the summary table~\ref{tab:astab}.
Similar analyses from experiments at the TRISTAN collider, around c.m. energies
of 58~GeV, gave \cite{as-tristan}
$$\as \left( 58\ {\rm GeV}\right) = 0.132\pm 0.008 \ ,$$
which is also included in the summary.

\begin{figure}[ht]
\begin{center}
\epsfxsize10.0cm\epsffile{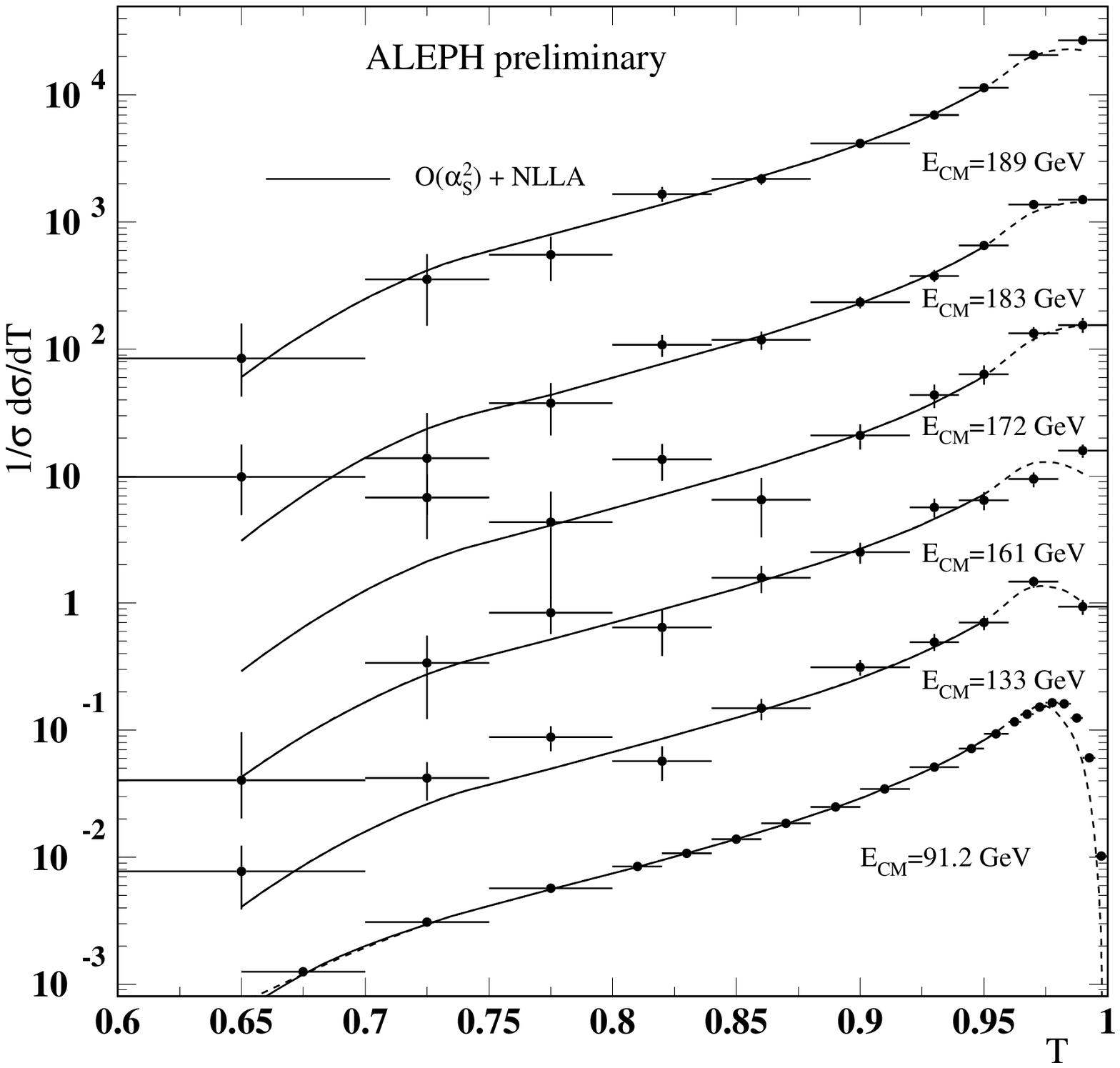} 
\end{center}
\caption{Differential Thrust distributions measured at various c.m.
energies at LEP (symbols), together with resummed NLO QCD fits (lines; figure
adapted from \cite{A-as}).}
\label{fig:a-shapes}
\end{figure}

At the LEP collider, all four experiments (ALEPH, DELPHI, L3 and OPAL) have
contributed a multitude of
studies based on the high statistics data samples around the $\z0$ resonance
(LEP-I; $\ecm \sim 91.2$~GeV) and at the higher energies of the LEP-II
running phase ($\ecm \sim 133,\ 161,\ 172,\ 183$ and 189~GeV).
The SLD experiment at the SLAC Linear Collider (SLC) contributed similar studies
at $\ecm \sim \mz$.
As an example of the precise description of data by the CQD fits,
the Thrust distributions as measured by ALEPH at LEP-I and LEP-II energies are
shown in figure~\ref{fig:a-shapes},
together with the corresponding fits of the resummed NLO QCD calculations.

The most current results from LEP and SLC, based on resummed NLO QCD
calculations, are taken from 
\cite{A-as,D-as,L-as,O-as,SLD-as} and are combined to obtain one single
value of $\as$ at each c.m. energy.
This is done by calculating weighted averages of the results quoted by each
experiment, with the inverse quadratic total error taken as the weight.
The experimental uncertainties are combined assuming that no correlation exists
between the experiments. 
Theoretical uncertainties are assumed to be common to all experiments,
however different methods and definitions were used to estimate these
in each case.
Therefore, a linear average of the theoretical uncertainties quoted by the
experiments is used to define the theoretical error of the combined results.
This gives
\begin{eqnarray}
\as \left( 91.2\ {\rm GeV}\right) &=& 0.121\pm 0.001 \pm 0.006\ , \nonumber \\
\as \left( 133\ {\rm GeV}\right) &=& 0.113\pm 0.003 \pm 0.006\ , \nonumber \\
\as \left( 161\ {\rm GeV}\right) &=& 0.109\pm 0.004 \pm 0.005\ , \nonumber \\
\as \left( 172\ {\rm GeV}\right) &=& 0.104\pm 0.004 \pm 0.005\ , \nonumber \\
\as \left( 183\ {\rm GeV}\right) &=& 0.109\pm 0.002 \pm 0.004\ , \nonumber \\
\as \left( 189\ {\rm GeV}\right) &=& 0.110\pm 0.001 \pm 0.004\ , \nonumber
\end{eqnarray}
where the first errors are experimental and the second theoretical.
These results are included in the final summary table~\ref{tab:astab}.
Note, however, that some of the averages at LEP-II energies contain
results which are still preliminary, see \cite{A-as,D-as,L-as,O-as}.
 
The energy dependence of $\as$ from studies of event shape observables is
clearly seen in the results presented above.
This is also demonstrated  in a dedicated analysis of the L3 Collaboration
\cite{L3-as-rad} which includes a study of radiative events recorded at LEP-I.
Such events are effectively produced at reduced hadronic centre of mass energies;
they can therefore be used to determine $\as$ at energy scales below the nominal
collider energy.
The statistical precision at these reduced energies is, however, limited.

\begin{figure}[ht]
\begin{center}
\epsfxsize12.0cm\epsffile{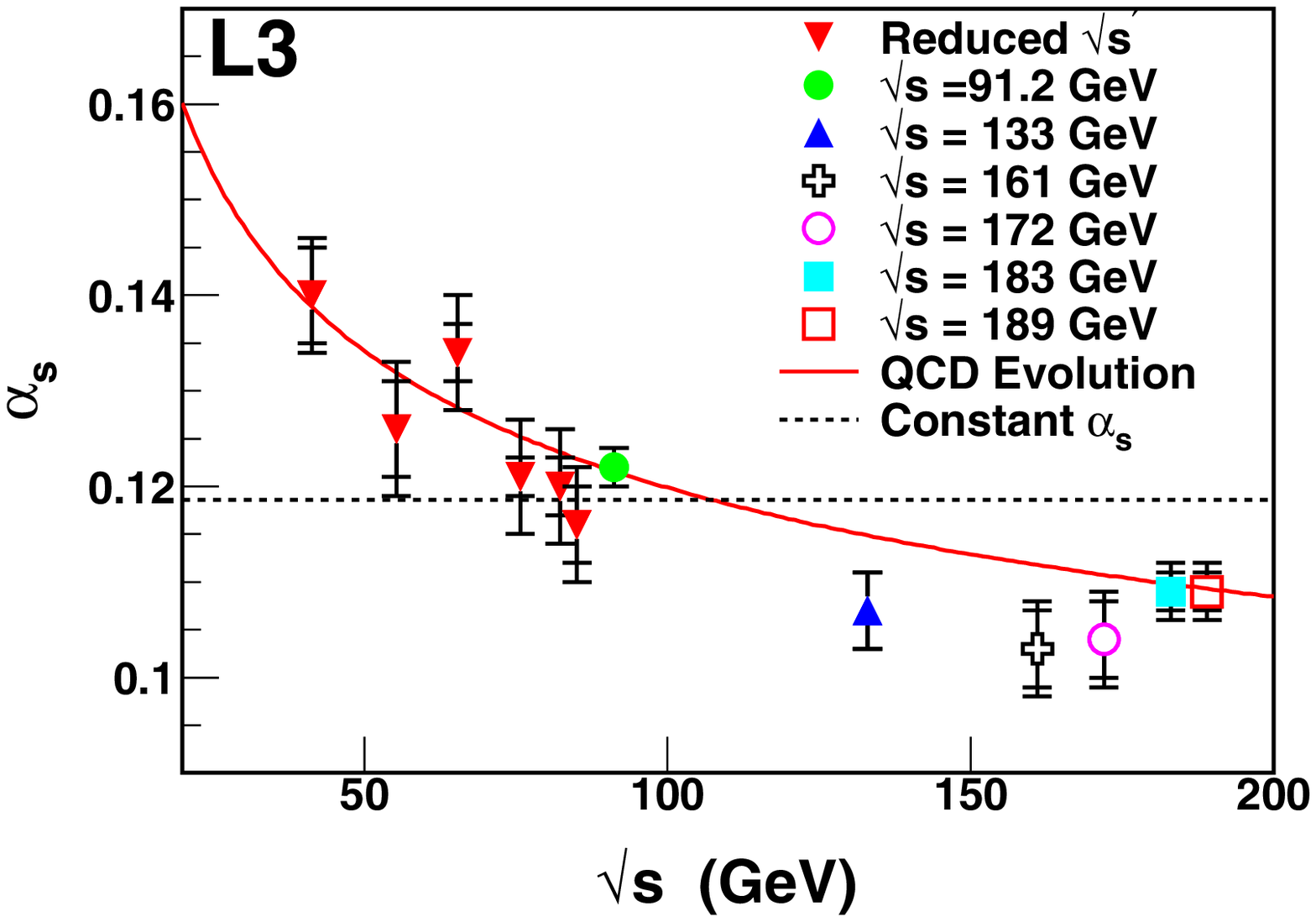}
\end{center} 
\caption{Running of $\as$ from hadronic event shapes at LEP, measured by L3.
The results at energies below 91~GeV are from radiative events 
at $2 E_{beam} \approx \mz$ (figure from reference~\cite{L3-as-rad}).}
\label{fig:l3-as}
\end{figure}

The results of the L3 study are shown in figure~\ref{fig:l3-as}, demonstrating the
agreement of the data with the QCD expectation of a running coupling.
Note that, in order to judge the significance for the running, only the
innermost, experimental errors must be taken into account, since theoretical
uncertainties are highly correlated between the different data points.
The running of $\as$ was also demonstrated in a recent study of jet production
rates at PETRA and at LEP energies, using the same consistent analysis method and
data from two experiments comprising similar detector techniques, JADE and OPAL
\cite{o-j-jets}.

Before the advent of resummed QCD calculations for event shape observables,
determinations of $\as$ were done in pure NLO, see e.g.
\cite{d-old-as,o-grand-as}.
Studies of theoretical uncertainties typically included scale variations from
$\xmu = 1$ down to the best fit values of $\xmu$, from two-parameter fits, or the
optimized scales given by the methods discussed in Section~2.7~.
These ranges were quite large, down to $\xmu \sim 0.01$, which lead to rather
large scale uncertainties in case of some of the observables.
Resummed NLO results, however, had smaller dependencies on scale variations,
as theoretically expected,
which is why they developed as a standard in $\as$ determinations from event
shape studies.

\subsubsection{Pure NLO results}

Fits of pure NLO QCD to experimental event shape and jet rate distributions
are known to provide a consistent description of the data if - in addition to
$\as$ - also the renormalization scale is treated as a free parameter of the fit
\cite{siggiscale}.
While the fitted scale factor $\xmu$ assumes different values for
different observables, indicating that the amount of unknown, higher order
contributions is different in each case, the resulting value of $\as$ appears to
be \oq universal" and in much better agreement than applying a single and fixed
choice of scale \cite{d-old-as,o-grand-as}.

These findings are corroborated in a
recent re-analysis of DELPHI data at LEP-I
\cite{D-scale2}, which also demonstrates that  experimentally optimized NLO
calculations can provide a more consistent description of the data than resummed
NLO at fixed scale $\xmu = 1$.
Defining a scale range, for each observable, of a factor of 2
around the experimental fit value of $\xmu^2$, results in $\amz = 0.1174 \pm
0.0026$, which includes both experimental and theoretical uncertainties.
This method of optimizing renormalization scales and defining the size of the
remaining theoretical uncertainties seems logical from an experimental point of
view, see the discussion in Section~2.7, but carries intrinsic theoretical
problems: 

Firstly, the resulting scale factors are rather small, down to
$\cal{O}$(0.01) or smaller, such that large logarithms appear in the NLO
coefficients, see equation~13.
Secondly, there is no common agreement about the significance of $one$ overall
scale for each distribution because theoretical scale optimizations predict
$\xmu$ to depend on the value of the observable itself.
The issue of experimental scale optimization therefore requires further
investigation and discussion.

\subsubsection{Power corrections} \nonumber

In the past few years, analytical approaches were
pursued to approximate nonperturbative hadronization effects by means of
perturbative methods, introducing a universal, non-perturbative parameter
$$\alpha_0 (\mu_I ) = \frac{1}{\mu_I } \int^{\mu_I}_0 {\rm d} k\ \as (k) $$
to parametrize the unknown behaviour of $\as (Q)$ below a certain
infrared matching scale $\mu_I $ \cite{powcor}.
Divergent soft gluon contributions to the perturbative predictions of event
shapes are removed and, as a consequence of these techniques, lead to
corrections which are proportional to powers of $1/Q$.
Power corrections are regarded as an alternative approach to describe
hadronization effects on event shape distributions, instead of using
phenomenological hadronization models as in the studies discussed in the previous
sections.

The energy dependence of event shape distributions and of their integrated
mean values, in the c.m. energy range from 14 to 183~GeV, were analysed and
compared with perturbative QCD calculations plus added power corrections
\cite{j-as,j-powcor,A-as,D-as}. 
Two-parameter fits of $\alpha_0$ and $\amz$ to the data provide a
consistent description of the shapes and the energy evolution of the data;
however the scatter between results from different shape observables is larger
than the typical uncertainties would suggest.
The non-perturbative parameter $\alpha_0$ turns out to be universal for all
studied observables within about 20\%, and values of $\amz$ seem to be close
to but systematically smaller than those from fits using conventional methods of
hadronization corrections \cite{j-powcor}.

Power corrections, as an analytical ansatz to compute nonperturbative QCD
corrections to event shape observables, offer a rather promising alternative to
the use of phenomenological hadronization models in determinations of $\as$;
however more experience, confidence and further developments are necessary before
actual fit results of $\as$ can supersede the results from \oq classical" 
event shape analyses described above. 

\subsection{$\as$ from scaling violations of fragmentation functions}

The total cross section of $\epem$ annihilations into charged hadrons $h$, $\epem
\rightarrow h + X$, can be factorized into a perturbative and a nonperturbative
regime, describing processes of hard gluon radiation and of hadronization,
respectively.
The perturbative part can be calculated in terms of so-called coefficient
functions, while the non-perturbative part is parametrized by phenomenological
fragmentations functions.
The separation of these two regimes is performed at an energy scale
$\mu_f$, the factorisation scale, which --- as in DIS --- enters as a
second, arbitrary energy scale, in addition to the renormalization scale $\mu$.
According to the factorization theorem,
the cross section can then be written
\cite{nason-webber}:
\begin{equation}\label{eq:scal}
\frac{{\rm d} \sigma}{{\rm d}x} (\epem \rightarrow h + X) = 
\sum_i \int^1_x \frac{{\rm d}z}{z}\ C_i \left(z, \as(\mu^2), \mu^2/Q^2 \right)\ 
D_i \left( x/z, \mu_f^2 \right)\ ,
\end{equation}
where $C_i$ are the coefficient functions for creating a parton with flavour $i$
and momentum fraction $z = p_{parton} / p_{beam}$, $D_i$ represent the
probability that parton $i$ fragments into a hadron $h$ with momentum fraction
$x/z$, and $x = p_{hadron} / p_{beam}$.

In lowest order, the coefficient functions $C_i$ are given by the electroweak
couplings for quarks; they vanish for gluons.
Higher order QCD corrections apply to the quark coefficient functions and 
lead to finite functions for gluons, too.
These corrections are known up to complete NLO \cite{nlo-co-fctns}, however only
the LO  corrections were available and used in currently
existing analyses.

The fragmentation functions $D_i$ are not given by perturbation theory; however
--- as in the case of parton densities and structure functions (c.f.
Section~3.1) --- their energy dependence is described by QCD via complicated
integro-differential equations and the DGLAP evolution functions
\cite{dglap,ap}.

%
%

The ALEPH \cite{A-scal} and the DELPHI \cite{D-scal} collaborations at LEP have
both published detailed analyses of the charged hadron fragmentation functions,
using data samples from the PETRA, PEP and LEP colliders spanning c.m. energy
ranges from 14 to 91.2~GeV.
At LEP, ALEPH and DELPHI extracted differential $x$ distributions separately
for  initial b-, c- and uds-quark events, and obtained the corresponding
gluon-jet distribution from tagged gluon jet event samples.
Assuming a certain parametrization of the \oq bare" fragmentation functions at a
\oq starting energy scale" $Q_0$, $\amz$, the free parameters of the starting
fragmentation functions for gluons and the different quark flavours, as well as
a parameter allowing for additional, nonperturbative power law
corrections, are determined in fits to all experimental $x$-distributions.

Combining both the ALEPH and the DELPHI results, which are based on comparable
data sets and analysis strategies, leads to
$$\amz = 0.125 ^{+ 0.006}_{- 0.007}\ {\rm (exp.)}\ \pm 0.009\ {\rm (theo.)}\,$$
where the theoretical uncertainty includes variations of both the factorization
and the renormalization scales\footnote{
DELPHI chose $\mu_f \equiv \mu$ and varied both within 1/2 to 2~$\ecm$; ALEPH
varied each independently within $1/\sqrt{e}$ to $\sqrt{e}~\ecm$ and added both
uncertainties in quadrature. 
Here, the uncertainties quoted by DELPHI are taken because they are less
restrictive in limiting the range of scales.}.

\subsection{$\as$ from total hadronic cross sections}

Because of its inclusive nature, the total cross section of the process $\epem 
\rightarrow$~hadrons was the first quantity for
which QCD corrections up to complete NNLO were known \cite{gorishny,chetyrkin}.
For c.m. energies in the $\epem$ continuum, far below the $\z0$ pole and for
$\mu \equiv \ecm$, the normalized cross section is given by
\begin{equation} \label{eq-Rgamma}
R_{\gamma} = \frac{\sigma (\epem \rightarrow {\rm hadrons})}{\sigma (\epem
\rightarrow \mu^+\mu^-)} = 3 \sum_i  q_i^2 \left( 1 + \frac{\as}{\pi} + 1.441
\left( \frac{\as}{\pi} \right)^2 - 12.8 \left( \frac{\as}{\pi}\right)^3\right)\ ,
\end{equation}
where $q_i$ are the electrical charges of quark flavours $i$ which are produced,
i.e. for which $2 M_q \ll \ecm$.

Because of the above functional form of $R_{\gamma}(\as)$, $relative$ errors of
$R_{\gamma}$ lead to $absolute$ errors in $\as$ of about the same size, $\Delta R /
R \sim \Delta \as$, such that precise measurements of $R$ still lead to rather
large errors in $\as$.
As another complication, electroweak corrections due to real $\z0$ exchange and due
to $\gamma - \z0$ interference must be applied to the data in the c.m. energy range
above 20 GeV.
These corrections are negligible at $\ecm$~=~14~GeV, but at $\ecm$~=46~GeV
they amount to about the same size as the QCD corrections \cite{cello-R}.

A combination of the cross section measurements of all four PETRA experiments, in
the c.m. energy range of 14~to~46.8~GeV, accounting for correlated
experimental uncertainties, electroweak corrections and leading order initial
state radiation effects,
resulted in $\as  {\rm (34\
GeV)}  = 0.169 \pm 0.025$ \cite{cello-R}, in NLO QCD.
Applying the full NNLO QCD prediction which was not yet available at that time
results in $\as  {\rm (34\ GeV)}  = 0.175 \pm 0.028$ or  ---
equivalently --- in $\amz = 0.143 \pm 0.018$.
Two groups have analysed and combined $\epem$ hadronic cross sections measured in
the c.m. energy ranges from 7~to 57~GeV \cite{R-dagostini} and from 2.65~to
52~GeV \cite{R-marshall}.
They both used an initially erroneous value of the NNLO QCD coefficient of
$R_\gamma$, +64.8 instead of -12.8; when corrected for this mistake, they result
in $\as (34\ {\rm GeV}) = 0.165 \pm 0.022$ \cite{R-dagostini} and in
$\as (31.6\ {\rm GeV}) = 0.158 \pm 0.019$ \cite{R-marshall}.
Combining the two gives
$ \as (34\ {\rm GeV}) = 0.160 \pm 0.019\ {\rm (exp. + sys.)}$ or 
$\amz = 0.133 \pm 0.013$.

A re-analysis \cite{haidt} of PETRA and TRISTAN data, in the c.m. energy range of
20 to 65~GeV, took better account of higher order QED and electroweak corrections
and used the mass of the $\z0$ measured from LEP experiments.
In NLO QCD, this study resulted in $\amz = 0.124 \pm 0.021$, where the error
includes experimental and systematical uncertainties, added in quadrature.
Re-applying the NNLO QCD correction this gives
$$ \as  {\rm (42.4\ GeV)} = 0.175 \pm 0.028\ \ {\rm or}\ \
\amz = 0.126 \pm 0.022\ ,$$
which is included in the final summary section~7.
A more recent determination of $\as$ from the total $\epem$ hadronic cross section
measured by CLEO at $\ecm = 10.52$~GeV \cite{cleo-rhad} resulted in
$$ \as  {\rm (10.52\ GeV)} = 0.130\ ^{+\ 0.021\ }_{-\ 0.029\ }$$
and is also added to the final summary.

On top and around the $\z0$ resonance, the LEP experiments have collected large
statistics data samples which allow accurate determinations of $\as$.
At the $\z0$ pole, equation~\ref{eq-Rgamma} must be modified according to the
dominant electroweak couplings of the $\z0$ to the quarks, resulting in NNLO QCD
predictions of the hadronic decay width of the $\z0$ \cite{rznnlo}.
Quark mass corrections in NLO \cite{xmasses} and partly to
NNLO \cite{xmasses2}, non-factorizable electroweak and QCD corrections
\cite{rzcorr-aas}, top quark effects and other electroweak corrections apply in
addition, rendering numerical expressions for 
$\rz = \Gamma (\z0 \rightarrow {\rm hadrons}) / \Gamma (\z0
\rightarrow {\rm leptons})$ rather complicated; see e.g. ref.~\cite{rz-report}
for a comprehensive report about QCD corrections on $R_{\gamma}$ and $\rz$.

In this review, a recent parametrisation of the NNLO QCD prediction of $\rz$
\cite{r-param}, including all known corrections indicated above, is applied to
determine $\as$ from $\rz$.
This parametrisation, as given in equation~\ref{eq-Rz-coeff}, 
is also used by the LEP collaborations in their combined
studies of electroweak precision data \cite{lep-ew}.
The coefficients given in equation~\ref{eq-Rz-coeff} are calculated
for a Higgs mass $M_H$ of 300~GeV,
a top quark mass $M_t$ of 174.1~GeV and for $\mz$~=~91.19~GeV.
With the latest combined LEP result,
$\rz = 20.768 \pm 0.024$ \cite{lep-ew}, this gives 
$ \amz = 0.124 \pm 0.004\ {\rm (exp.)}$.

\renewcommand{\arraystretch}{1.3}
\begin{table}[h]
\caption{Estimates of errors in the determination of $\amz$ from $\rz$, caused
by different sources of uncertainties.
\label{tab:r-err}}
\begin{center}
\begin{tabular}{|l|l|}
   \hline 
error source & $\Delta \amz$  \\
\hline 
$\Delta \mz = \pm 0.0021\ {\rm GeV} $ & $ \pm 0.00003 $ \\
$\Delta M_t = \pm 5\ {\rm GeV} $ & $ \pm 0.0002 $ \\
$M_H = 100\ ...\ 1000\ {\rm GeV} $ & $ \pm 0.0017 $\\
$\mu = \left( \frac{1}{4}\ ...\ 4\right)\ \mz $ & $ ^{+\ 0.0028}_{-\
0.0004} $\\
renormalization schemes & $\pm 0.0002$ \\
\hline
total & $ ^{+\ 0.003}_{-\ 0.002}$ \\
\hline
\end{tabular}
\end{center}
\end{table}

Errors in $\amz$ from different sources, namely from changes of $\mz$ and
$M_t$ within their current uncertainties, from $M_H$ in the range of 100~GeV (the
current experimental lower limit from LEP) to 1000~GeV, and from renormalization
scale uncertainties, varying $\mu$ from 1/4 to 4$\mz$, are given in
Table~\ref{tab:r-err}.
The estimate of the $scheme$ dependence, which in NNLO must be considered
in addition to the renormalization $scale$ dependence, was taken from reference
\cite{kataev-starshenko}. 
Neglecting the tiny errors from $\mz$ and from $M_t$, the overall theoretical
uncertainty on $\amz$ from $\rz$ is estimated to be
$\pm 0.002$ from $M_H$ and $M_t$ and $^{+\ 0.003}_{-\ 0.001}$ from higher order
QCD contributions.
Therefore, the final result from $\rz$, to be added to the overall summary, is
$$ \amz = 0.124 \pm 0.004 {\rm (exp.)} \pm 0.002 (M_H, M_t ) ^{+\ 0.003}_{-\
0.001} {\rm (QCD)}\ . $$
Note that the precision of this result crucially depends on the assumption that 
the predictions of the electroweak Standard Model are strictly valid; small
deviations from these predictions can produce large systematic shifts of $\amz$
from $\rz$. 
The LEP data, however, are in excellent overall agreement with the
Standard Model predictions \cite{lep-ew}, such that the uncertainties quoted
above seem realistic.

A combined fit to all data from LEP-I and LEP-II, including the measurement of
the mass of the W boson, $M_W$, and all measured cross sections and asymmetries,
instead of $\rz$ alone, results in
$\amz = 0.120 \pm 0.003 {\rm (exp.)}$; a fit to all data including those from
$p\overline{p}$ collider and lepton-nucleon scattering experiments gives
$\amz = 0.118 \pm 0.003 {\rm (exp.)}$ \cite{lep-ew}.
These fits include simultaneous determinations of $M_H$ and $M_t$, such that only
the QCD uncertainty must be added.
Because of their complicated nature, however, these results are only mentioned for
completeness, but are not taken to replace the above value of $\amz$ from $\rz$
for the final summary.

\subsection{$\as$ from $\tau$ decays}

An important quantity to determine $\as$ from measurements at small energy scales
is the normalized hadronic branching fraction of $\tau$ lepton decays,
\begin{equation}
R_{\tau} = \frac{\Gamma (\tau \rightarrow {\rm hadrons}\ \nu_{\tau})} {\Gamma (\tau
\rightarrow {\rm e} \nu_e \nu_{\tau})}\ ,
\end{equation}
which is predicted to be~\cite{braaten}
\begin{equation} R_\tau = 3.058
( 1.001 + \delta_{\rm pert} + \delta_{\rm nonpert})\ .
\end{equation}
Here, $\delta_{\rm pert}$ and $\delta_{\rm nonpert}$ are perturbative and
nonperturbative QCD corrections; $\delta_{\rm pert}$ was calculated to
complete $\oaaa$ and is of similar structure to the one for $\rz$
\cite{gorishny,braaten,lediberder92}:
\begin{equation}
\delta_{\rm pert} =
\frac{\as (M_{\tau})}{\pi} + 5.20
\left( \frac{\as (M_{\tau})}{\pi} \right)^2 +26.37 \left(
\frac{\as (M_{\tau})}{\pi}\right)^3\ ,
\end{equation}
and parts of the $4^{th}$ order coefficient are also known \cite{tau-4}.
Based on the operator product expansion (OPE) \cite{ope},
the nonperturbative correction was estimated to be small~\cite{braaten},
$\delta_{\rm nonpert} = -0.007 \pm 0.004$.

The most comprehensive determinations of $\as$ from $\tau$ decays are based on
recent studies from LEP, making use of the large data statistics available at
LEP-I.
The ALEPH \cite{A-tau} and the OPAL \cite{O-tau} Collaborations presented
measurements of the vector and the axial-vector contributions to the differential
hadronic mass distributions of $\tau$ decays, which allow 
simultaneous determination of
$\as$ and of the nonperturbative corrections (in terms of the OPE).
These corrections were found to be small and to largely cancel in the total sum of
$R_{\tau}$, in good agreement with the theoretical estimates.

\renewcommand{\arraystretch}{1.3}
\begin{table}[h]
\caption{Results of $\as (M_{\tau})$ from $R_{\tau}$, for different variants of
NNLO QCD calculations.
\label{tab:astau}}
\begin{center}
\begin{tabular}{|l|l c c|l c c|}
   \hline 
  & \multicolumn{3}{c|} {ALEPH \cite{A-tau}} 
  & \multicolumn{3}{c|} {OPAL \cite{O-tau}}   \\ 
Theory & $\as (M_{\tau})$ & exp. & theo. & $\as (M_{\tau})$ &exp. & theo. \\
\hline \hline 
CIPT & 0.345 & $\pm 0.007$ & $\pm 0.017$ & 0.348 & $\pm 0.010$ & $\pm 0.019$ \\
FOPT & 0.322 & $\pm 0.005$ & $\pm 0.019$ & 0.324 & $\pm 0.006$ & $\pm 0.013$ \\
RCFT & \ --- &             &             & 0.306 & $\pm 0.005$ & $\pm 0.011$ \\
\hline
\end{tabular}
\end{center}
\end{table}

The final results of $\as (M_{\tau})$ are listed in Table~\ref{tab:astau},
obtained for different variants of the NNLO QCD predictions: fixed order
perturbation theory (FOPT) \cite{braaten}, contour improved perturbation theory
(CIPT) \cite{tau-4}, expressing $\delta_{\rm pert}$ by contour  integrals in the
complex $s$-plane, and renormalon chain improved perturbation theory (RCPT)
\cite{rcpt}, where leading terms of the $\beta$-functions are
resummed by inserting so-called renormalon chains,
i.e. gluon lines with many loop insertions.
The two groups agree well on their $\as$ results and --- approximately --- on the
estimated uncertainties, however different theoretical approaches give systematic
differences in $\as$.
The FOPT results seem to represent the mean of these theoretical approaches.
Therefore the final result from $R_{\tau}$, to be included in the final summary
section~7, is taken to be
$$
\as (M_{\tau}) = 0.323 \pm 0.005 {\rm (exp.)} \pm 0.030 {\rm (theo.)}\ , 
$$
where the average between ALEPH and OPAL was taken and an additional
error of $\pm 0.020$, accommodating the shift between different theoretical
approaches, was added in quadrature\footnote{In another study of the impact of
different theoretical variants of the NNLO QCD expectation for $R_{\tau}$, it was
concluded that the overall theoretical uncertainty on $\as (M_{\tau})$ is, at best,
$\pm 0.05$
\cite{neubert-tau}, which is larger than the respective error quoted above.} 
to the theoretical uncertainties given by
ALEPH and OPAL.
When extrapolated to the energy scale $\mz$, using the 4-loop $\beta$-function
and 3-loop matching at the bottom quark pole mass, this results
in $\amz = 0.1181 \pm 0.0007 {\rm (exp.)} \pm 0.0030 {\rm (theo.)}$.

\section{Results from hadron colliders} \label{sec:ppbar}

Significant determinations of $\as$ from hadron collider data are obtained from
$b\overline{b}$ production cross sections, from
prompt photon production, from inclusive jet production cross sections and from
the ratio of $W+(1-jet)$ and $W+(0-jet)$ production cross section, all of which are
calculated in complete NLO QCD.
The latter topic, $W+jet$ production, will not be discussed further here,
because early measurements \cite{w+jets}, giving $\as (M_W) = 0.123 \pm
0.025$, were put in question by new analyses showing bad disagreement between
data and QCD, see e.g. \cite{w-jets}.

Hard scattering cross sections initiated by two hadrons with four-momenta $P_1$
and $P_2$ can be written as
\begin{equation} \label{eq-hadron-xsec}
\sigma (P_1,P_2) = \sum_{i,j} \int {\rm d}x_1 {\rm d}x_2 f_i (x_1,\mu_f^2) f_j
(x_2,\mu_f^2) \cdot  \sigma_{ij} \left( p_1,p_2,\as (\mu^2),Q^2/\mu_f^2\right) \ ,
\end{equation}
where $p_1 = x_1 P_1$ and $p_2 = x_2 P_2$ are the momenta of the interacting
partons, $f_i$ and $f_j$ are the QCD quark and gluon distributions, $Q$ is the
characteristic scale of the hard scattering, and $\sigma_{ij}$ is the
short-distance cross section of the hard scattering between partons of type $i$
and $j$.
This parametrization again is based --- as in DIS --- on the
assumption of factorization between the short- and the long-range regimes of the
scattering process, the transition between both being defined at the
factorization scale $\mu_f$.

In general, determinations of $\as$ from hadron-hadron-collisions are less
precise than those from $\epem$ annihilation or deep inelastic scattering
processes, due to larger uncertainties associated with incoming hadrons:
parton distributions and soft remnants from spectator partons, which do
not participate in the hard scattering process, substantially add to the overall
uncertainties.

\subsection{$\as$ from $b\overline{b}$ cross sections}

The first theoretically well-defined determination of $\as$ from a purely
hadronic production process was
presented by the UA1 collaboration~\cite{ua1-bb}, obtained from a measurement of
the cross section of the process
$p \overline{p} \rightarrow b \overline{b} X$ for
which NLO QCD predictions exist \cite{bb-theo}.
$b$-quarks were detected through their semileptonic decays into muons which yield
high transverse momenta $p_T$ with respect to the beam axis.
A strong correlation between the decay muons and the original $b$-quark allows
determination of the $b$-quark production cross section without application of a
jet algorithm, thus avoiding systematic effects from spectator partons (or the \oq
underlying event").
The LO QCD contribution is of $\oaa$ and leads to a back-to-back 
(in azimuth) $b\overline{b}$
configuration; virtual corrections to this state and the emission of a third
(gluon) jet are of $\oaaa$, predicted in NLO QCD.

Comparison of the measured cross-section for 2-body final states with NLO QCD
predictions yielded 
$$ 
\as (20\ {\rm GeV}) = 0.145^{+\ 0.012}_{-\ 0.010}\ {\rm (exp.)} ^{+\ 0.013}_{-\
0.016}\ {\rm (theo.)}\ ,
$$
where the theoretical error includes uncertainties due to
different sets of structure functions, renormalization/factorization scale
uncertainties and the b-quark mass~\cite{ua1-bb}.

\subsection{$\as$ from prompt photon production}

Production of high transverse momentum direct (\oq prompt") photons in hadron
collisions is well suited to test perturbative QCD and to determine $\as$,
because photons, in contrast to quarks, do not hadronize and their energies and
directions can --- in general --- be measured with higher accuracy than those of
hadron jets.
In leading order, however,
prompt photon production is of $\cal{O}(\alpha \as)$, compared to $\oaa$ for
hadron jets, and therefore suffers from relatively small production cross
sections.
In addition, there is a sizeable background of photons from $\pi^0$ and $\eta$
decays, such that quantitative tests of QCD from prompt photon production are not
trivial, from an experimental point of view.

Using complete ${\cal O}(\alpha \as^2)$ QCD predictions
\cite{prompt-theo}, the UA6 collaboration determined $\as$ from a measurement of
the cross sections difference 
$\sigma (p\overline{p} \rightarrow \gamma X) - \sigma (pp \rightarrow \gamma X)$
\cite{ppgam-recent}, where the poorly known contributions of the sea quarks and the
gluon distributions in the proton cancel. 
The result is
$$
\as (24.3\ {\rm GeV}) = 0.135 \pm 0.006\ ({\rm exp.})
^{+\ 0.011}_{-\ 0.005}\ ({\rm theo.})\ ,
$$
where the theoretical error includes uncertainties from the scale choice and from
variation of the parton distribution functions.

\subsection{$\as$ from inclusive jet cross sections}

The definition and reconstruction of particle jets in hadron
collisions traditionally has followed other strategies than those used in
$\epem$ annihilation.
In hadron collisions, so-called cone jet finders are employed which allow
particles, ideally those which originate from the proton remnants, not to be
associated with any of the reconstructed jets --- in contrast to the clustering
algorithms used in $\epem$ annihilation where $all$ particles are assigned to
jets; c.f. section~4.1.
Nowadays, almost all of the jet studies at hadron collider experiments follow the
\oq Snowmass" definition of jets \cite{snowmass-jet}.
Here, jets are defined by concentrations of transverse energy $E_T = | E \sin
\theta |$ in cones of radius
$$
R = \sqrt{(\Delta \eta)^2 + (\delta \phi)^2}\ ,
$$
where $\eta = - \ln \tan (\theta /2)$ is the pseudorapidity, $\phi$ is the
azimuthal and $\theta$ is the polar angle of a particle or an energy cluster
in the calorimeter of the detector, measured w.r.t. the point of beam crossing.

\begin{figure}[ht]
\begin{center}
\epsfxsize12.0cm\epsffile{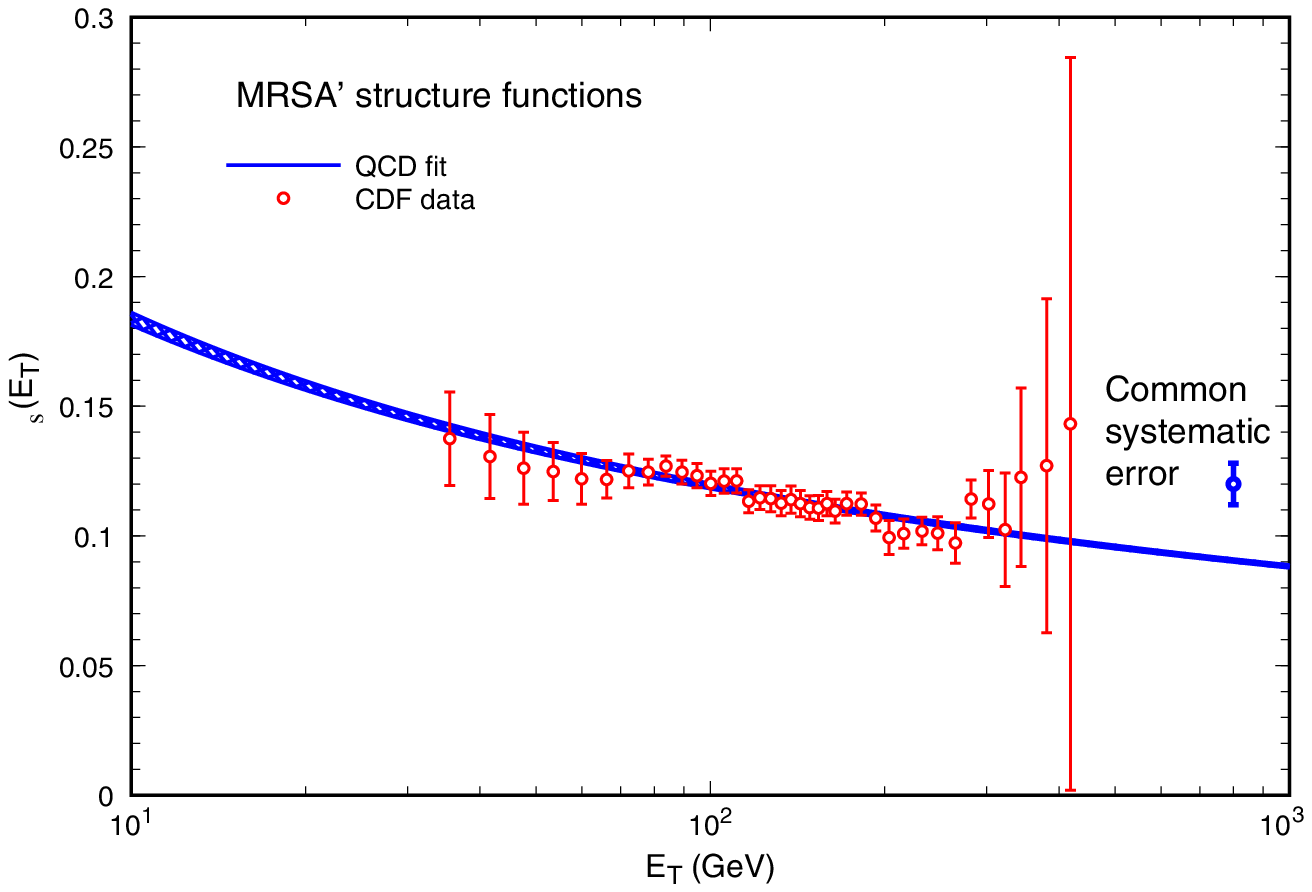}
\end{center} 
\caption{Values of $\as (E_T)$ extracted from CDF inclusive jet cross section
data (figure adapted from \cite{ppjet}).}
\label{fig:giele-jets}
\end{figure}

Giele, Glover and Yu determined $\as$ \cite{ppjet} by fitting the NLO parton
level Monte Carlo JETRAD \cite{jetrad}, which is based on the QCD matrix elements
of reference~\cite{qq-jet-me} and the MRSA' parton density functions \cite{mrsda},
to the single inclusive jet cross sections measured by CDF \cite{cdf-jet} for a
cone size of $R = 0.7$.
In each bin of $E_T$, $\as$ is determined for the scale choice  $\mu \equiv
E_T$. 
The resulting values of $\as (E_T)$ are displayed in figure~\ref{fig:giele-jets},
where the error bars represent the combined statistical and theoretical
uncertainties; the size of an additional, overall (experimental) systematic error
is also indicated.
The results are in good agreement with the QCD expectation of the running $\as$.
A corresponding overall QCD fit, in NLO, gives
$$\amz = 0.121 \pm 0.001\ {\rm (stat.)} \pm 0.008\ {\rm (sys.)} \pm 0.005\ {\rm
(theo.)}\ \pm 0.002\ {\rm (pdf)}\,
$$
where the theoretical error was obtained from a variation of $\mu$ between 0.5
and $2 \cdot E_T$, and the last error represents the uncertainty from using
different parton density functions.

This study was basically meant to introduce a general method and possibility to
determine $\as$ from hadron collider jet cross sections, rather than to present a
complete and final analysis.
For instance, the parton density functions used in this study were extracted from
(mainly deep inelastic scattering) data for a fixed input value of $\amz = 0.113$,
while a coherent determination of $\as$ should allow $\as$ to vary in the density
functions, too.
Although the data statistics --- and hopefully also the experimental systematic
errors --- must have improved significantly with respect to the previous data
sample of reference~\cite{cdf-jet}, and although there exist new parametrizations
of density functions for different input values of $\as$, no update of the 
measurement of $\as$ from hadron collider jets was published so far.
Therefore the above result is retained in the final summary --- keeping in
mind, however, that the quoted uncertainties constitute lower
limits rather than a complete assessment of the overall error.

\section{Results from heavy quarkonia decays and masses} \label{sec:qqbar}


The mass spectra and partial decay widths of heavy quark-antiquark bound states 
are a good testing ground for QCD.
For quark masses $m_Q \gg \Lambda_{QCD}$, the short- and long-range effects on the
decay widths can be factorized, and the short-range part can be calculated by
perturbative QCD.
Mass splittings of heavy quarkonia states can be calculated using nonperturbative
methods like lattice gauge theory, which also provide means to extract values of
$\as$.

\subsection{$\as$ from quarkonia decay branching fractions}

Partial decay widths of heavy quarkonia, like $\Gamma^{\mu\mu} = \Gamma
(Q\overline{Q} \rightarrow \mu^+ \mu^-)$, $\Gamma^{\gamma g g}$ and $\Gamma^{ggg}$
are calculated in NLO QCD \cite{hq-nlo}. 
The expressions contain the (unknown) radial
wave function at the origin, which can be eliminated by forming the ratios
\begin{eqnarray}
R_{\mu} &\equiv& \frac{\Gamma^{ggg}}{\Gamma^{\mu \mu}} = \frac{10 (\pi^2
-9)}{9\pi} 
  \frac{\as^3(\mu^2)}{\alpha^2} \left( 1 + \left[ 0.4 - 6.3 \ln \left(
\frac{m_b^2}{\mu^2} \right) \right] \frac{\as (\mu^2)}{\pi} \right) \nonumber \\
R_{\gamma} &\equiv& \frac{\Gamma^{\gamma gg}}{\Gamma^{ggg}} = \frac{4}{5}
  \frac{\alpha}{\as (\mu^2)} \left( 1 - \left[ 2.6 - 2.1 \ln \left(
\frac{m_b^2}{\mu^2} \right) \right] \frac{\as (\mu^2)}{\pi} \right)\ .
\end{eqnarray}

An early, comprehensive fit of $\as$ from the $J / \Psi$ and the $\Upsilon$ 
branching ratios was presented in reference~\cite{kobel}, resulting in
$\amz = 0.113^{+0.007}_{-0.005}$.
This study included estimates of the higher order QCD uncertainties and of
relativistic corrections.

More recently, sum rules for the $\Upsilon$ system were analysed with resummation
of QCD Coulomb effects which are responsible for the growth of the perturbative
coefficients in NLO QCD \cite{Ydec-recent}, resulting in $\amz = 0.118 \pm 0.006$.
Similar methods were already applied in ealier studies \cite{hq-early} but have led
to conflicting results. 
Application of NNLO QCD predictions resulted in \cite{Ydec-3rd}
$$\amz = 0.118 \pm 0.006\ ,$$
which is included in the final summary of $\as$.
It should be noted that 
further studies of $\Upsilon$ sum
rules in NNLO QCD \cite{hoang} obtained similar results of $\as$,
however with a much more conservative estimate of
the theoretical uncertainties --- the resulting, large uncertainties of $\as$ may
indicate that $\Upsilon$ sum rules do not appear to provide competitive results
of $\as$.

\subsection{$\as$ from lattice calculations of mass splittings}

Early determinations of $\as$ from heavy quarkonia mass splittings were based on
quenched approximations of lattice QCD calculations,
neglecting light quark flavour loops ($N_f = 0$). 
When evolved to the $\msbar$ coupling at the scale $\mz$, these methods led to
$\amz = 0.105 \pm 0.004$ \cite{lgt-pion}, where the error
included statistical as well as estimates of systematic errors.

Refined nonrelativistic QCD calculations with $N_f=0$ and $N_f=2$
allowed extrapolation to the physically correct number of light quarks, $N_f = 3$,
and included advanced (3-loop) perturbative extrapolation from the lattice to the
$\msbar$ coupling.
A detailed study of various mass splittings of $\Upsilon$ ($b\overline{b}$)
states, which are precisely known from corresponding measurements, was presented
in  \cite{lgt-davies}.
Technically, the lattice calculation for a given value of the \oq bare" lattice
coupling $\alpha_{lat}$ gave a value for the dimensionless quantity $a \Sigma$,
where
$a$ is the lattice spacing and $\Sigma$ is the mass splitting of
suitable quarkonium states. 
From this result, divided by the measured value of $\Sigma$, the lattice
spacing $a$ was obtained.
From QCD perturbation theory, the $\msbar$ coupling $\as$ at the energy scale
$a^{-1}$ is given as a function of $\alpha_{lat}$ \cite{lgt-msbar}.
Extraction of $\as ( a^{-1} )$ and extrapolation to $\amz$ resulted
in $\amz = 0.1174 \pm 0.0024$ \cite{lgt-davies}, where the error included
statistical as well as systematic uncertainties which include
variations of light quark masses and estimates of truncation errors in the
extraction of the $\msbar$ coupling using perturbation theory.

More recently, a study based on similar lattice calculations and the same
$\Upsilon$ mass splittings as in reference~\cite{lgt-davies}, however using a
different discretization scheme, derived a lower value, $\amz = 0.1118 \pm
0.0017$ \cite{lgt-spitz}. 
This is about three standard deviations smaller than
the result from reference~\cite{lgt-davies},
which led to the conclusion that the \oq true" systematic uncertainty
must be three to four times larger than estimated before.

Following this suggestion, the result on $\as$ from nonperturbative lattice
calculations to be considered in the final summary section is chosen to
be the average of the two results discussed above, with the overall
uncertainty increased by a factor of 3, giving
$$ \amz = 0.115 \pm 0.006.$$

\section{Summing up ...} \label{sec:summary}

A summary of the $\as$ measurements 
discussed in the previous sections is presented in table~\ref{tab:astab}
at the end of this review.
The results are given, if applicable, at the relevant energy scale $Q$ of the
process, $\as (Q)$, and at the standard \oq reference" scale of the $\z0$ mass,
$\amz$. The conversion between these two cases was done \cite{as4} using the
4-loop QCD expression for the running $\as$, equation~\ref{eq-as4loop}, with
3-loop matching at the c- and b-quark pole masses of 1.5 and 4.7~GeV,
respectively, as discussed in section~2.5.
The splitting of the overall uncertainties of $\amz$, $\Delta \amz$, into
experimental and theoretical errors is given in the $5^{th}$ and the $6^{th}$
columns of table~\ref{tab:astab}, and the last column indicates the level of
theoretical calculations on which these results are based.

\begin{figure}[ht]
\begin{center}
\epsfysize12.0cm\epsffile{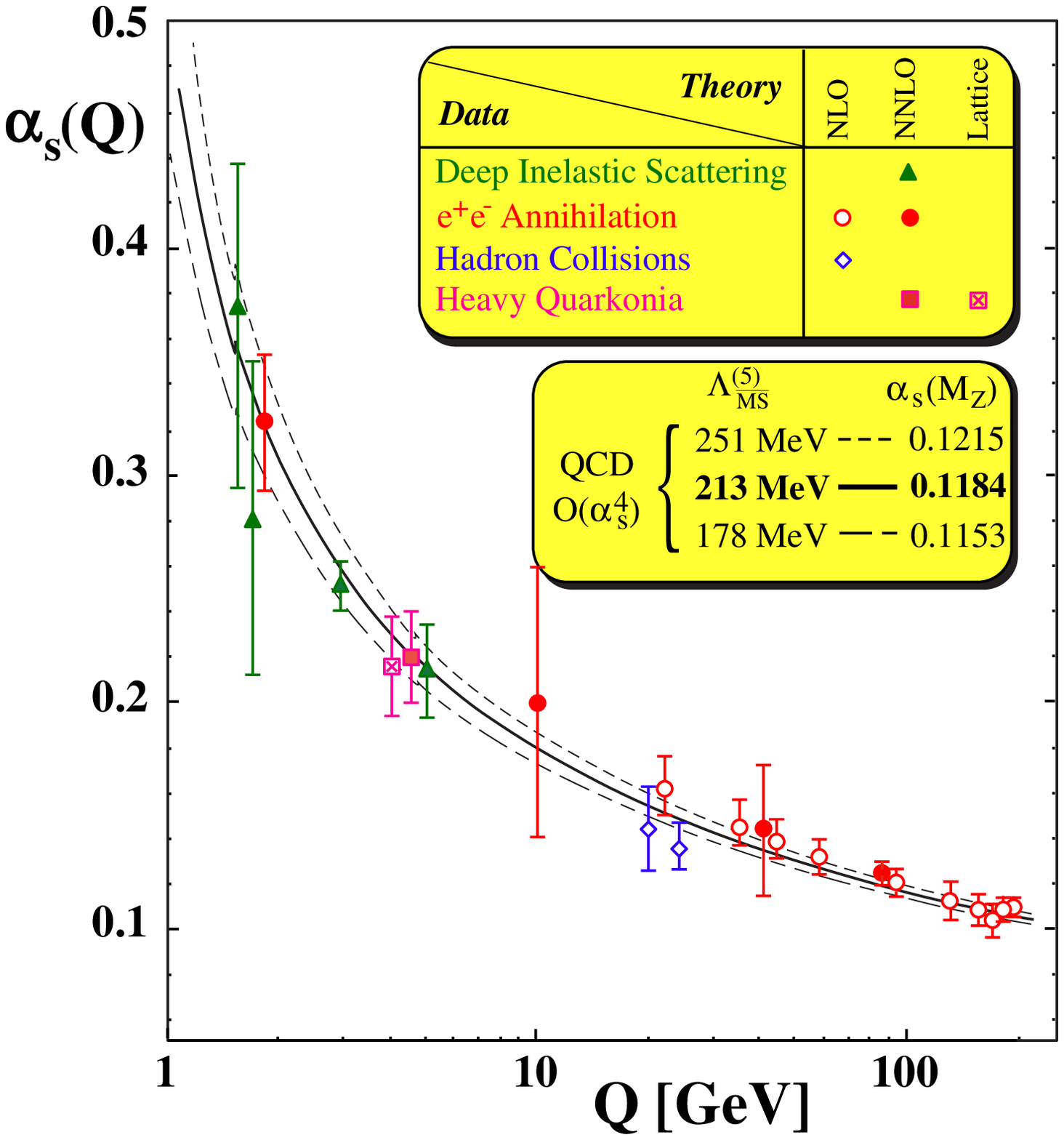} 
\end{center}
\caption{Summary of $\as (Q)$ .
\label{fig:as-q}}
\end{figure}

The results for $\as (Q)$, given in the $3^{rd}$ column of table~\ref{tab:astab},
are presented in figure~\ref{fig:as-q}, together with fits of the 4-loop QCD
prediction for the running $\as$ (equation~\ref{eq-as4loop}) with 3-loop matching
at the quark pole masses.
Results which were obtained from data in large ranges of $Q$ are
not displayed in this figure.
The data are in very good agreement with the theoretical expectation, and
prove the running of $\as$ with high significance.
The latter point will be analysed in more
detail in section~7.2.

\begin{figure}[ht]
\begin{center}
\epsfysize12.0cm\epsffile{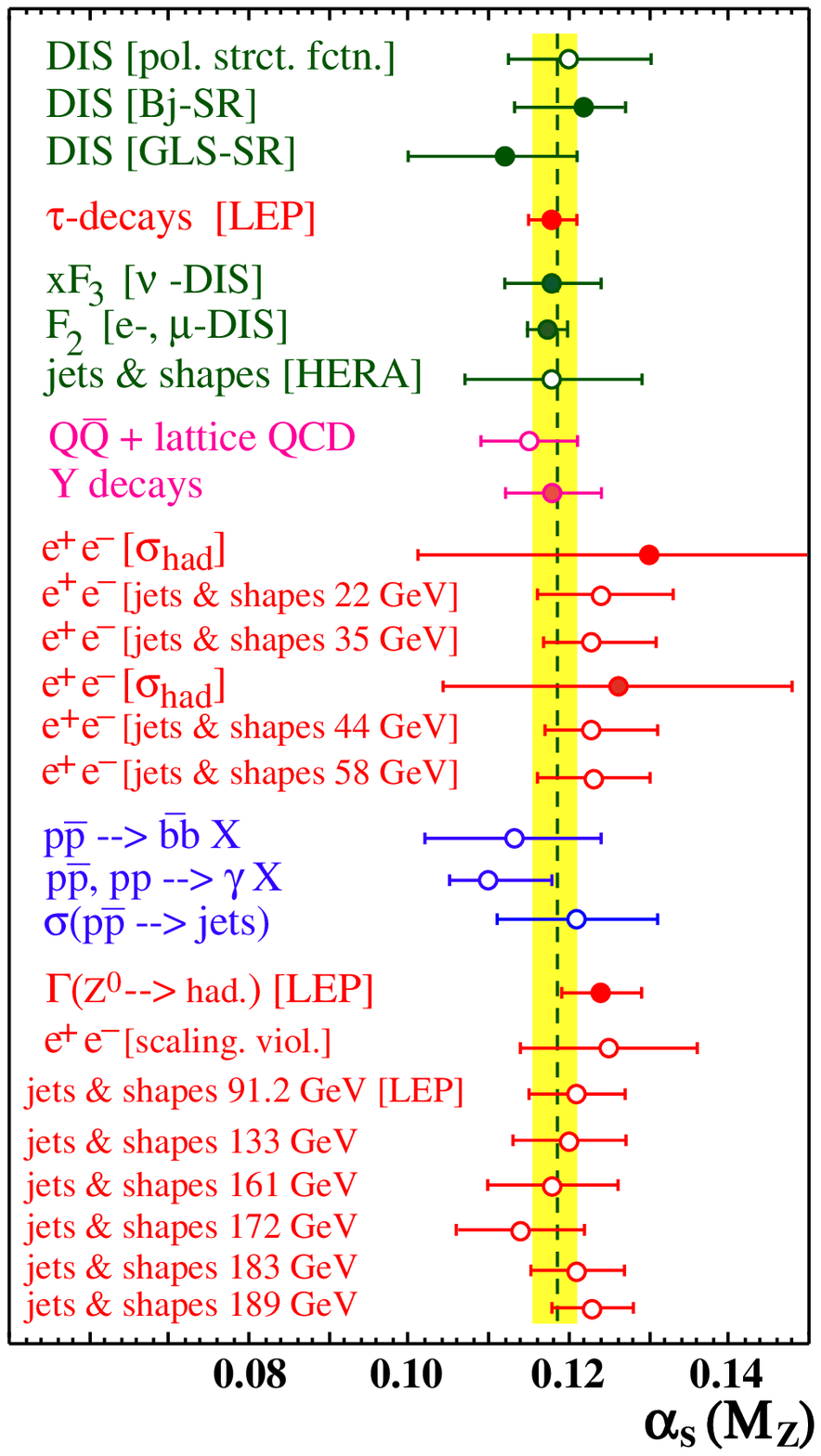} 
\end{center}
\caption{Summary of $\amz$ (filled symbols represent results based on complete
NNLO QCD).
\label{fig:as-mz}}
\end{figure}

The values of $\amz$ are presented in figure~\ref{fig:as-mz}.
Within their assigned total erros, all results agree well with each other and
with a weighted average value of $\amz = 0.1184$, which is indicated by the 
vertical line in figure~\ref{fig:as-mz}.
The overall $\chi^2$ of all results with this average is 7.2
for 25 degrees of freedom, which indicates that the individual errors must be
highly correlated. 
Therefore, the determination of the average value of $\amz$ and its overall
uncertainty requires special treatment and discussion.

\subsection{World average $\wamz$ and its overall uncertainty $\dwas$}

The errors of most $\as$ results are dominated by theoretical uncertainties,
which are estimated using a variety of different methods and definitions.
The significance of these nongaussian errors is largely unclear.
Furthermore, there are large correlations between different results, due to
common theoretical uncertainties, as e.g. for event shape
measurements in $\epem$ annihilations.
Correlations between
$\as$ determinations from different processes, such as DIS and $\epem$
annihilations, or between different procedures and observables used within the
same class of processes may be present, too.
Standard statistical methods therefore do not apply when averaging these
results.
Several methods are employed to derive an estimate of the average value $\wamz$
and its overall uncertainty, $\dwas$.
The results are summarized in
table~\ref{tab:aserr}:

\begin{itemize}
\item
An error weighted average and an \oq optimized
correlation" error is calculated from the error covariance matrix, assuming
an overall correlation factor between the total errors of all measurements.
This factor is adjusted so that the overall $\chi^2$ equals one per degree
of freedom~\cite{schmelling}.
The resulting mean values, overall uncertainties and optimized
correlation factors are given in columns 3 to 5 of
Table~\ref{tab:aserr}, respectively.
\item
For illustrative purposes only, an overall error is calculated assuming that all
measurements are entirely uncorrelated and all quoted errors are gaussian.
The results are displayed in column 6.
\item
The simple, unweighted root mean squared of the mean values of all
measurements is calculated  and shown in column 7, labelled \oq simple
rms".
\item
Assuming that each result of $\amz$ has a rectangular-shaped rather than a
gaussian probability
distribution, 
the resulting weights (the inverse of the square of the total
error) are summed up in a histogram, and the resulting $rms$ of that distribution
is quoted as \oq rms box" \cite{qcd97}.
\end{itemize}

All of these methods have certain advantages but also include inherent
problems.
The \oq simple rms" indicates the scatter of all results around
their common mean, but does not depend on the individual errors
quoted for each measurement. 
The \oq box rms", which takes account of the errors
and of their nongaussian nature, was criticized as being
too conservative an estimate of the overall uncertainty of $\as$. 
The \oq optimized correlation" method --- in the absence of a detailed knowledge of
these correlations --- over-simplifies by the  assumption of one
overall correlation factor. 
Moreover, if correlations are present,
$\chi^2$  does not have the same mathematical and probabilistic meaning as in the
case of uncorrelated data. 
In the extreme, $\chi^2$ may even be negative.

\renewcommand{\arraystretch}{1.3}
\begin{table}[htb]
\caption{
Average values of $\wamz$ and averaged uncertainties, for several
methods to estimate the latter, and for several subsamples 
of the available data. \label{tab:aserr} }
\begin{center}
  {
\begin{tabular}{|c|l|c|c|c||c|c|c|}
   \hline
& & & opt. corr. & overall & uncorrel. & simple rms &  rms box \\
row & sample \hfill (entries)& $\wamz$ & $\dwas$ & correl. &
  $\dwas$ & $\dwas$ 
  & $\dwas$ \\
\hline
1 & all \hfill (26)         & 0.1191 & 0.0045 & 0.71 & 0.0012 & 0.0043 & 0.0057\\
2 & \ $\Delta \as\le 0.010$ \hfill (20)
                        & 0.1191 & 0.0041 & 0.66 & 0.0012 & 0.0037 & 0.0051\\
3 & \ $\Delta \as\le 0.008$ \hfill (18)
                        & 0.1190 & 0.0039 & 0.62 & 0.0012 & 0.0038 & 0.0050\\
4 & \ $\Delta \as\le 0.006$ \hfill (9)
                        & 0.1188 & 0.0033 & 0.64 & 0.0014 & 0.0029 & 0.0038\\
5 & \ $\Delta \as\le 0.005$ \hfill (4)
                        & 0.1189 & 0.0022 & 0.28 & 0.0017 & 0.0034 & 0.0033\\
 & & & & & & &\\
6 & NNLO only \hfill (9)    & 0.1185 & 0.0035 & 0.78 & 0.0016 & 0.0045 & 0.0048\\
7 & \ $\Delta \as\le 0.008$ \hfill (6)
                        & 0.1184 & 0.0031 & 0.68 & 0.0016 & 0.0026 & 0.0032\\
8 & \ $\Delta \as\le 0.005$ \hfill (3)
                        & 0.1184 & 0.0022 & 0.27 & 0.0018 & 0.0037 & 0.0028\\
9 & \ $\Delta \as\le 0.004$ \hfill (2)
                        & 0.1175 & 0.0026 & 0.95 & 0.0019 & 0.0006 & 0.0019\\
&  & & & & & &\\
10 & only DIS\hfill (6)      & 0.1178 & 0.0040 & 0.94 & 0.0020 & 0.0014 & 0.0047\\
11 & only $\epem$\hfill (15) & 0.1209 & 0.0051 & 0.79 & 0.0016 & 0.0038 & 0.0054\\
12 & only $p\overline{p}$  \hfill (3)  
                        & 0.1135 & 0.0074 & 0.60 & 0.0051 & 0.0059 & 0.0068\\
 & & & & & & &\\
13 & $Q \le 10$~GeV  \hfill (9)
                        & 0.1177 & 0.0040 & 0.93 & 0.0016 & 0.0017 & 0.0042\\
14 & $10 < \frac{Q}{{\rm GeV}} < 90$ \hfill (9)
                        & 0.1202 & 0.0064 & 0.56 & 0.0029 & 0.0062 & 0.0077\\
15 & $Q \ge 90$~GeV  \hfill (8) 
                        & 0.1213 & 0.0056 & 0.78 & 0.0023 & 0.0035 & 0.0050\\
\hline
\end{tabular} }
\end{center} 
\end{table}

\begin{figure}[ht]
\begin{center}
\epsfxsize13.5cm\epsffile{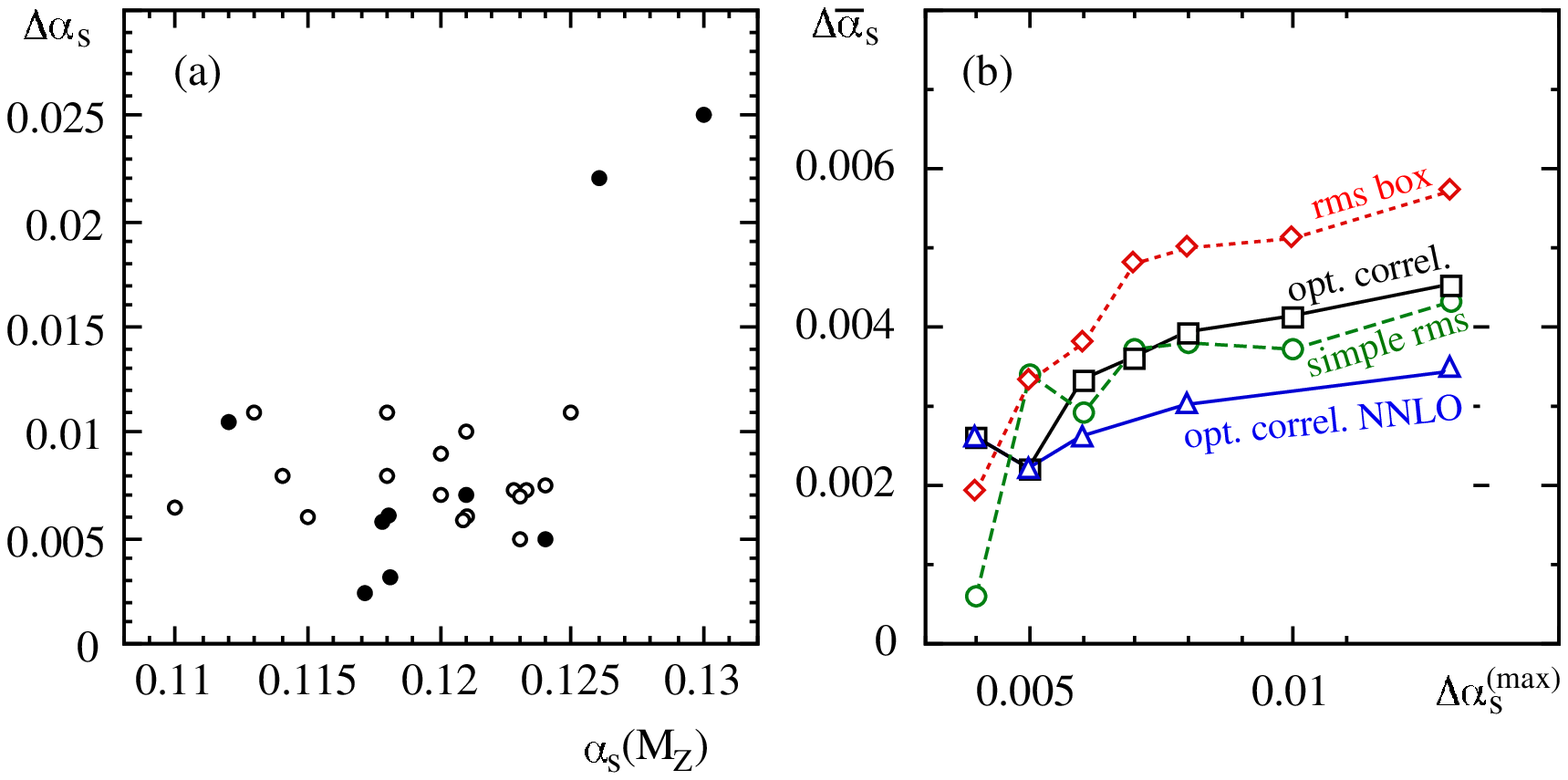}
\end{center} 
\caption{(a) Results of $\amz$ versus their respective overall errors; full
symbols are based on complete NNLO QCD.
(b) Dependence of $\dwas$ on the
selection of results with errors $\Delta\amz \le \Delta\as^{(max)}$, 
for different methods to calculate $\wamz$.
The rightmost data points correspond to no cut on $\Delta\as^{(max)}$.}
\label{fig:as-scat}
\end{figure}

With these reservations in mind, all four methods do provide some  estimate of 
$\dwas$. 
Apart from the method to calculate $\dwas$, the result also depends on
the significance of the data included in the averaging process: in all cases
except the \oq uncorrelated" error estimate,
$\dwas$ is largest if all data are included, and
tends to smaller values if the averaging is restricted to results
with errors $\Delta\as \le \Delta\as^{(max)}$, i.e. if only the most
significant results are taken into account.
This is demonstrated in rows 1 to 5 of table~\ref{tab:aserr}.
The dependence of $\dwas$ on $\Delta\as^{(max)}$ is illustrated in
figure~\ref{fig:as-scat}b, where the rightmost results correspond to the
first row of table~\ref{tab:aserr} which includes all 26
$\as$ measurements summarized in table~\ref{tab:astab}.
Figure~\ref{fig:as-scat}a illustrates the distribution of $\amz$ and the total
errors of all these measurements.

On first sight it seems logical to restrict the determination of $\wamz$, and
especially of $\dwas$, to the most significant data, if the inclusion of 
less significant measurements only enlarges $\dwas$.
Taken to the extreme, one may even be tempted to take the one result with
the smallest quoted error as the final world average value of $\wamz$.
However, the errors on $\amz$ estimated in individual
studies are very often {\em lower limits} because unknown
and additional systematic effects can only increase the total error.
In some cases, 
small total errors can be due to 
fortunate coincidences, to ignorance,
over-optimism and/or neglect of certain error sources.
In order to ensure and to test consistency of the results,
it is therefore mandatory not to rely on a single determination alone but
to possibly include several different, significant measurements
in the averaging process.

The averaging procedures discussed so far include results which are based on
different orders and types of QCD calculations, namely on lattice gauge theory
and on perturbation theory in NLO, resummed NLO and in NNLO.
Comparing and averaging these different types of results is nevertheless
justified because all of them include estimates of the relevant
uncertainties, such that they should be compatible with a
common average within their given errors. 
In order to achieve the highest precision and confidence in a combined average
value $\wamz$, however, it may be beneficial not to include
results which are, for instance, based on calculations in lower than the maximum
available order of perturbation theory.

In this sense, the averaging procedure is applied only to those results which
are based on complete NNLO QCD perturbation theory.
Only very recently the number, the precision and the diversity of such
measurements reached a level where such a restriction still provides a solid
basis for a meaningful averaging process which includes sufficient freedom for
internal consistency checks.
Table~\ref{tab:astab} contains 9 measurements which are based on
NNLO QCD; the results of averaging these are given in rows 6 to 9 of
table~\ref{tab:aserr}, for different subsamples defined by selecting
measurements with a total error $\Delta \as \le \Delta \as^{(max)}$.
The combined errors, $\dwas$, using the optimized correlation method, are
displayed in figure~\ref{fig:as-scat}b.

Requiring $\Delta \as \le 0.008$ rejects those $\as$ measurements which are
dominated by large experimental uncertainties, leaving the 5 most significant NNLO
results from which one obtains, using the optimized correlation method,
$$
\wamz = 0.1184 \pm 0.0031\ .
$$ 
This value is taken as the currently best estimate of the world average
of $\amz$.
According to 
equations~\ref{eq-as4loop} and~\ref{Mq-matching}, in 4-loop approximation and
with 3-loop threshold matching at $M_b = 4.7$~GeV, 
this corresponds to
$\lamsb^{(N_f=5)} = \left( 213^{+38}_{-35} \right)$~MeV and
$\lamsb^{(N_f=4)} = \left( 296^{+46}_{-44} \right)$~MeV.

Each of the 26 single results summarized in table~\ref{tab:astab} is
compatible with this average, to within about one standard deviation of its
assigned uncertainty or less. 
In order to investigate possible systematic deviations or trends between and
within subsets of these measurements, the averaging procedure was repeated for
deep inelastic scattering data, for $\epem$ annihilation and for hadron collider
data alone, see rows 10 to 12 of table~\ref{tab:aserr}, and for 3 energy ranges
as shown in rows 13 to 15. 
Within the overall uncertainties (e.g. those
derived from the optimized correlation method), all results
agree well with each other and with the world average value derived above. 
Small but insignificant systematic differences between DIS and
$\epem$ results and between those obtained from low and from high energy data
may be visible; these may
well be accidental or may be caused, for example, by different methods of treating
renormalization and factorization scales.

\subsection{Quantifying the running of $\as$: determination of $\beta_0$, of $N_c$
and of the functional $Q$-dependence }

The precision of the experimental results, the large energy range for which data
are available, and the good agreement of the measurements with the QCD
expectation of a running $\as$, see figure~\ref{fig:as-q}, suggest actually
fitting the functional form of the energy dependence from the data.
For this purpose, a subset of data is selected which ensures maximal
independence between the data chosen and which is based on the most significant
measurements in the largest possible energy range (c.f. table~\ref{tab:astab} and 
figure~\ref{fig:as-q}), namely:
\begin{itemize}
\item $\as$ from $\tau$ decays:\hfill $\as (1.778\ {\rm GeV}) = 0.323 \pm 0.030$;
\item $\as$ from moments of $F_2$:\hfill $\as (2.96\ {\rm GeV}) = 0.252 \pm 0.011$;
\item $\as$ from scaling violations of $F_3$:\hfill $\as (5.0\ {\rm GeV}) = 0.214
\pm 0.021$;
\item $\as$ from the hadronic width of the $\z0$:\hfill $\as (91.2\ {\rm GeV}) =
0.124
\pm 0.005$;
\item  $\as$ from hadronic event shapes:\hfill $\as (189\ {\rm GeV}) = 0.110 \pm
0.004$.
\end{itemize}
All these results are based on complete NNLO QCD predictions --- with the exception
of the latter which includes resummed NLO calculations --- and represent the most
precise $\as$ results of their class.
The nature of their experimental and theoretical uncertainties ensures a maximum
of independence between them.

\renewcommand{\arraystretch}{1.3}
\begin{table}[htb]
\caption{
Functional fits of $\asq$, in the energy range of 
$1.778~{\rm GeV} \le Q \le 189~{\rm GeV}$.
The LO QCD expectation for $B \equiv 1/\beta_0$, with $N_c = 3$ and $N_f = 5$, is
1.64. }\label{tab:asfits}
\begin{center}
\begin{tabular}{|l|c|c|c|c|c|}
   \hline
$\as (Q) =...$ & A & B & C & $\chi^2$ / d.o.f. & prob. \\
\hline
$B \left( \ln \frac{Q^2}{C^2} \right)^{-1}$ &
    --- & $1.62 \pm 0.09$ & $0.125 \pm 0.032$ & 0.63 / 3 & 0.89 \\
  & & & & & \\
$A + C * Q$ & $0.201 \pm 0.007$ & --- & $(-5.3 \pm 0.5)\cdot 10^{-4}$ & 78 / 3
& $10^{-16}$ \\
$A + C / Q$ & $0.113 \pm 0.003$ & --- & $0.41 \pm 0.03$ & 4.6 / 3 & 0.20\\
\hline
\end{tabular} 
\end{center} 
\end{table}

Table~\ref{tab:asfits} summarizes some of the functional fits which were
performed with these selected data.
The general idea was to obtain qualitative measures for the functional form of
the energy dependence of $\as$, as well as numeric fit values for the coefficient
$\beta_0$ of the QCD $\beta$-function, see equation~\ref{eq-betafunction}, and
eventually for one of the main parameters of QCD, the number of colour degrees of
freedom $N_c \equiv C_A$ which is an integral part of $\beta_0$:
\begin{equation}\label{eq-beta0}
\beta_0 = \frac{11 C_A - 2 N_f}{12 \pi}\ .
\end{equation}
Because of its simplicity and in order to avoid too many QCD-inspired biases, 
the functional form of the leading order QCD expression
for the running $\as$ ( c.f. equation~\ref{eq-as4loop}), without quark threshold
matching, was fitted to the data.  
Higher order corrections to the
running affect $\as$ at the smallest energy scales, $Q \sim M_{\tau}$, by less
than $10\%$, which is well within the error of $\as$ in this energy regime.

The results of this fit are given in the first row of table~\ref{tab:asfits},
with $B \equiv 1/\beta_0 = 1.62 \pm 0.09$, and $C \equiv \Lambda = 0.125 \pm
0.032$ (in units of GeV).
The $\chi^2$ of this fit is 0.63 for 3 degrees of freedom, which corresponds to
a probability of 0.89.
Within QCD, with $N_c = 3$ and $N_f = 5$, $1/\beta_0 = 1.64$, which is
in excellent agreement with the fit value of the parameter $B$.
$\Lambda = 0.125$~GeV corresponds, in LO QCD, to $\amz = 0.124$ and to $\as
(M_{\tau}) = 0.31$ (without threshold matching and with $N_f=5$ throughout).
From the fit value of $B \equiv 1 / \beta_0$ one can derive, according to
equation~\ref{eq-beta0} and for $N_f = 5$ quark flavours,
$$ N_c = 3.03 \pm 0.12\ ,$$
which is an excellent and precise verification of the QCD group structure which
implies $N_c = 3$.
Alternatively, because this functional fit actually constrains the {\em ratio}
$B/n$, where $n$ is the power of $Q/C$ inside the logarithm, the fit determines,
when
$B$ is fixed to its QCD value of 1.64, that $n = 2.03\pm 0.12$.

Other functional forms of the energy dependence of $\as$, which are not predicted
by any consistent theory but which are added to demonstrate the significance of
the {\it logarithmic} decrease of $\as (Q^2)$, are fitted and presented in the last
two rows of table~\ref{tab:asfits}:
a straight-line fit is clearly excluded by the unacceptable $\chi^2$ of 78 for 
3~d.o.f., while a $1/Q$ energy dependence is disfavoured but cannot be excluded
with the current data.

\section{Conclusions} \label{sec:conclusion}

This review of experimental determinations of 
the coupling parameter of the Strong Interaction, $\as$,
summarized a topical and still ongoing field of activities in the high energy
physics community.
Due to impressive theoretical developments and experimental efforts during the
past 10 years, $\as$ could be determined from a large variety of physical
observables and processes with steadily increasing precision.
Current state-of-the-art measurements reach experimental uncertainties down to a
few per cent, and theoretical calculations which are complete up to NNLO
perturbative QCD obey higher order uncertainties of a few per cent
in $\amz$, too. 
While not all of the available studies reach both these levels of
precision, the most significant determinations of $\as$, based on complete
NNLO QCD calculations, can be summarized to a new world average value of
$$
\wamz = 0.1184 \pm 0.0031\ .
$$
The overall uncertainty is slightly larger than the smallest total
errors quoted for some of the measurements, which is due to the following facts:
\begin{itemize}
\item
The errors of single results contain different estimates
of theoretical uncertainties, which are not uniquely defined.
\item
Many (if not all) results rely on assumptions which ultimately cannot be proven,
like factorization between perturbative and nonperturbative effects, the nature
and size of the latter, neglect of quark masses, exact realization of the
electro-weak theory including the existence of a Standard Model Higgs boson,
reaching of the continuum limit in lattice calculations etc.; the quoted
theoretical uncertainties must therefore rather be viewed as lower limits instead
of complete estimates.
\item 
Many of the results and their theoretical uncertainties are highly correlated,
however mostly to an unknown degree.
This applies not only to their quoted errors, but also to the unproven assumptions
mentioned under the previous item.
\end{itemize}
A realistic determination of the world average value of $\as$ must therefore
be based on comparing several different measurements of similar precision.
It must account for possible correlations between them and allow for
underestimates and fortunate cancellations of errors in some cases.
In the absence of any \oq exact" method to account for all these effects, the
\oq optimized correlation" method \cite{schmelling} was used to calculate the
world average $\wamz$ and its overall remaining uncertainty as given above.
The total error depends on the averaging method chosen - alternative procedures
or preselections of data resulted in values ranging from $\pm 0.0022$ to $\pm
0.0057$.

All values of $\as$ summarized in table~\ref{tab:astab} agree well, within
their quoted uncertainties, with the world average.
Averages obtained from various subsamples of these data, like those from deep
inelastic scattering, from
$\epem$ annihilations and from hadron colliders, as well as subsamples from
different energy ranges, do not show significant biases or shifts.
In fact, the measured energy dependence 
of $\asq$ is in excellent agreement with the QCD expectation and
significantly proves the running of $\as$.

Functional fits to data in the energy range from 1.78 to 189~GeV
provide evidence for a logarithmic (in contrast to e.g. a linear) energy
dependence, and the fitted  logarithmic slope --- if interpreted in
terms of $\beta_0$, the LO coefficient of the QCD $\beta$-function ---  provides an
accurate value for the number of colour degrees of freedom, $N_c = 3.03 \pm 0.12$.
This is not actually a {\it measurement} of $N_c$, because many of the
$\as$ determinations are based on theoretical predictions of observables which
inherently include the QCD value of $N_c = 3$. 
Nevertheless, the fit result of
$N_c$ from the measured energy dependence of $\as$ constitutes an important
consistency check between data and QCD.

The total uncertainty of $\wamz$ quoted above is $2.6\%$.
This precision is a
remarkable  success. 
The error on $\wamz$, however, is much larger than those on other fundamental \oq
constants" of nature, like the fine structure constant $\alpha$, the weak
mixing angle $\sin^2 \theta_W$ or the gravitational constant \cite{pdg}.

Any further reduction of the uncertainties on $\as$ will require large
theoretical as well as experimental efforts:
\begin{itemize}
\item
New and higher order QCD calculations at least for some observables will be
mandatory to understand, specify and possibly reduce theoretical
uncertainties; most wishful candidates would be complete NNLO calculations for
jet rates and hadronic event shapes in $\epem$ annihilation and in DIS, which are
well understood experimentally.
\item
The precision and theoretical understanding of parton density functions and
their correlations with extracted values of $\as$ must be significantly improved.
Future high statistics and high energy runs of the Tevatron hadron collider
will hopefully provide this important input.
New theoretical developments to understand and predict structure functions from
basic principles, for instance through application of classical string theory,
see e.g. reference~\cite{string-sf}, are mandatory.
\item
So far, the neglect of finite (heavy) quark masses in almost all higher order QCD
calculations potentially limits the reliability especially of results
based on data which are close to
these thresholds.
While phase-space effects can be experimentally examined, only complete higher
order QCD calculations for massive quarks, see e.g. reference~\cite{qmasses}, can
close this gap of potentially large systematic uncertainties.
\item
The r\^ole of nonperturbative, long distance effects and the application of
factorization in many processes and $\as$ determinations must be further
investigated and understood.
Power corrections replacing or supplementing hadronization models, see e.g.
references~\cite{powcor,j-powcor}, or higher twist
corrections from lattice QCD, see e.g. reference~\cite{lat-ht}, are promising new
developments which have the potential for new, significant insights and
improvements.
\end{itemize}

The physics of hadronic interactions at high energy colliders is rich and
colourful. 
Much has been achieved in the past, however there are still many
fundamental and open questions. 
Further and more precise determinations of the strong coupling
parameter $\as$ will continue to concern and employ many motivated scientists in
the future.

\section*{Acknowledgments}
I am grateful to O. Biebel, A. Hoang, A.L. Kataev, J.H. K\"uhn, B. Webber and P.
Weisz for many interesting discussions, suggestions and comments.

\renewcommand{\arraystretch}{1.2}
\begin{table}[h,t]
{
\caption{
World summary of measurements of $\as$
(DIS = deep inelastic scattering; GLS-SR = Gross-Llewellyn-Smith sum rule;
Bj-SR = Bjorken sum rule;
(N)NLO = (next-to-)next-to-leading order perturbation theory;
LGT = lattice gauge theory;
resum = resummed NLO). \label{tab:astab}}
\begin{center}
\begin{tabular}{|l|c|c|c|c c|c|}
   \hline 
  & Q & & &  \multicolumn{2}{c|}
{$\Delta \amz $} &  \\ 
Process & [GeV] & $\alpha_s(Q)$ &
  $ \amz$ & exp. & theor. & Theory \\
\hline \hline 
DIS [pol. strct. fctn.] & 0.7 - 8 & & $0.120\ ^{+\ 0.010}
  _{-\ 0.008}$ & $^{+0.004}_{-0.005}$ & $^{+0.009}_{-0.006}$ & NLO \\
DIS [Bj-SR] & 1.58
  & $0.375\ ^{+\ 0.062}_{-\ 0.081}$ & $0.121\ ^{+\ 0.005}_{-\ 0.009}$ & 
  -- & -- & NNLO \\
DIS [GLS-SR] & 1.73
  & $0.280\ ^{+\ 0.070}_{-\ 0.068}$ & $0.112\ ^{+\ 0.009}_{-\ 0.012}$ & 
  $^{+0.008}_{-0.010}$ & $0.005$ & NNLO \\
$\tau$-decays 
  & 1.78 & $0.323 \pm 0.030$ & $0.1181 \pm 0.0031$
  & 0.0007 &  0.0030 & NNLO \\
DIS [$\nu$; $x{\rm F_3}$]  & 5.0
  & $0.214 \pm 0.021$
   & $0.118\pm 0.006$   &
    $ 0.005 $ & $ 0.003$ & NNLO \\
DIS [e/$\mu$; ${\rm F_2}$]
     & 2.96 & $0.252 \pm 0.011$ & $0.1172 \pm 0.0024$ & $ 0.0017$ &
     $ 0.0017$ & NNLO \\
DIS [e-p; jets]
     & 6 - 100 &  & $0.118 \pm 0.011$ & $ 0.002$ &
     $0.011 $ & NLO \\
${\rm Q\overline{Q}}$ states
     & 4.1 & $0.216 \pm 0.022$ & $0.115 \pm 0.006 $ & 0.000 & 0.006
     & LGT \\
$\Upsilon$ decays
     & 4.75 & $0.22 \pm 0.02$ & $0.118 \pm 0.006
     $ & -- & -- & NNLO \\
$\epem$ [$\sigma_{\rm had}$] 
     & 10.52 & $0.20\ \pm 0.06 $ & $0.130\ ^{+\ 0.021\ }_{-\ 0.029\ }$
     & $\ ^{+\ 0.021\ }_{-\ 0.029\ }$ & 0.002 & NNLO \\
$\epem$ [jets \& shapes]  & 22.0 & $0.161\ ^{+\ 0.016}_{-\ 0.011}$ &
   $0.124\ ^{+\ 0.009}_{-\ 0.006}$ &  0.005 & $^{+0.008}_{-0.003}$
   & resum \\
$\epem$ [jets \& shapes] & 35.0 & $ 0.145\ ^{+\ 0.012}_{-\ 0.007}$ &
   $0.123\ ^{+\ 0.008}_{-\ 0.006}$ &  0.002 & $^{+0.008}_{-0.005}$
   & resum \\
$\epem$ [$\sigma_{\rm had}$]  & 42.4 &
 $0.144 \pm 0.029$ &
   $0.126 \pm 0.022$ & $0.022
   $ & 0.002 & NNLO \\
$\epem$ [jets \& shapes] & 44.0 & $ 0.139\ ^{+\ 0.011}_{-\ 0.008}$ &
   $0.123\ ^{+\ 0.008}_{-\ 0.006}$ & 0.003 & $^{+0.007}_{-0.005}$
   & resum \\
$\epem$ [jets \& shapes]  & 58.0 & $0.132\pm 0.008$ &
   $0.123 \pm 0.007$ & 0.003 & 0.007 & resum \\
$\p\bar{\p} \rightarrow {\rm b\bar{b}X}$
    & 20.0 & $0.145\ ^{+\ 0.018\ }_{-\ 0.019\ }$ & $0.113 \pm 0.011$ 
    & $^{+\ 0.007}_{-\ 0.006}$ & $^{+\ 0.008}_{-\ 0.009}$ & NLO \\
${\rm p\bar{p},\ pp \rightarrow \gamma X}$  & 24.3 & $0.135
 \ ^{+\ 0.012}_{-\ 0.008}$ &
  $0.110\ ^{+\ 0.008\ }_{-\ 0.005\ }$ & 0.004 &
  $^{+\ 0.007}_{-\ 0.003}$ & NLO \\
${\sigma (\rm p\bar{p} \rightarrow\  jets)}$  & 30 - 500 &  &
  $0.121\pm 0.010$ & 0.008 & 0.005 & NLO \\
$\epem$ [$\Gamma (\z0 \rightarrow {\rm had.})$]
    & 91.2 & $0.124\pm 0.005$ & 
$0.124\pm 0.005$ &
   $ 0.004$ & $^{+0.003}_{-0.002}$ & NNLO \\
$\epem$ scaling viol. & 14 - 91.2 &  & $0.125 \pm 0.011$ & 
  $^{+\ 0.006}_{-\ 0.007}$ & 0.009 & NLO \\
$\epem$ [jets \& shapes] &
    91.2 & $0.121 \pm 0.006$ & $0.121 \pm 0.006$ & $ 0.001$ & $
0.006$ & resum \\
$\epem$ [jets \& shapes]  & 133.0 & $0.113\pm 0.008$ &
   $0.120 \pm 0.007$ & 0.003 & 0.006 & resum \\
$\epem$ [jets \& shapes]  & 161.0 & $0.109\pm 0.007$ &
   $0.118 \pm 0.008$ & 0.005 & 0.006 & resum \\
$\epem$ [jets \& shapes]  & 172.0 & $0.104\pm 0.007$ &
   $0.114 \pm 0.008$ & 0.005 & 0.006 & resum \\
$\epem$ [jets \& shapes]  & 183.0 & $0.109\pm 0.005$ &
   $0.121 \pm 0.006$ & 0.002 & 0.005 & resum \\
$\epem$ [jets \& shapes] & 189.0 & $0.110\pm 0.004$ &
   $0.123 \pm 0.005$ & 0.001 & 0.005 & resum \\
\hline
\end{tabular}
\end{center}
}
\end{table}

\end{document}